# Non resonant transmission modelling with Statistical modal Energy distribution Analysis

L. Maxit[*], K. Ege, N. Totaro, J.L. Guyader

Laboratoire Vibrations Acoustique, INSA-Lyon,
25 bis, av. Jean Capelle, F-69621 Villeurbanne Cedex, France

Statistical modal Energy distribution Analysis (SmEdA) can be used as an alternative to Statistical Energy Analysis for describing subsystems with low modal overlap. In its original form, SmEdA predicts the power flow exchanged between the resonant modes of different subsystems. In the case of sound transmission through a thin structure, it is well-known that the non resonant response of the structure plays a significant role in transmission below the critical frequency. In this paper, we present an extension of SmEdA that takes into account the contributions of the non resonant modes of a thin structure. The dual modal formulation (DMF) is used to describe the behaviour of two acoustic cavities separated by a thin structure, with prior knowledge of the modal basis of each subsystem. Condensation in the DMF equations is achieved on the amplitudes of the non resonant modes and a new coupling scheme between the resonant modes of the three subsystems is obtained after several simplifications. We show that the contribution of the non resonant panel mode results in coupling the cavity modes of stiffness type, characterised by the mode shapes of both the cavities and the structure. Comparisons with reference results demonstrate that the present approach can take into account the non resonant contributions of the structure in the evaluation of the transmission loss.

## 1. Introduction

The airborne noise transmission of panels has been studied extensively in the past. The standard wave approach [1] gives a good estimation of sound transmission loss for simple infinite extended structures like single and double panels made of an isotropic homogenous material. The advantages of this method are: (a) it can be applied to a wide frequency range with a short computation time; and (b) it takes dominant physical phenomena into account like the mass law effect of the panel below the critical frequency and the resonant frequencies for multi-panel systems. Statistical Energy Analysis has also been used to evaluate the sound transmission of simple/double panels separating two rooms [2, 3]. A limitation of these two approaches lies in the fact that they cannot be easily applied to evaluate sound transmission through nonhomogenous or stiffened panels like the firewall of an automobile. Moreover, they are not adapted to evaluate the influence on transmission loss of excitation conditions such as source room geometry, source location, or the position of the panel in the source room [4].

Different studies have been performed to improve knowledge on the sound transmission of panels with different levels of complexity. The interaction between a panel and a cavity has been studied in detail by numerous authors [5-8]. Low-frequency airborne sound transmission through a single partition was studied using a modal approach in [9-11]. The influence of the geometry and the dimensions of the room-wall-room system on low-frequency sound transmission was highlighted. Moreover, it was shown that resonant modes in the room are the cause of frequency dependent variations of sound insulation at low frequency. The finite element approach was also used to investigate the room-panel-room system at low frequency

---

[12]. The expansion of the solution on a functional basis to describe structural-acoustic problems and increase the frequency band of investigation was presented in [13]. The Patch Transfer Function approach [14] was used to evaluate the sound insulation of a finite size double wall partition [4] and the influence of room characteristics on acoustic transmission was again pointed out. The sound transmission of stiffened panels for the naval and railway sectors was studied in [15-18]. In [18] the authors used the space harmonic expansion method to show that the spacing and mechanical properties of stiffeners can significantly influence the acoustic performance of stiffened panels. Tools based on the same approach were developed in [19] for predicting the sound transmission through honeycomb panels. More recently, the effects of finite dimensions on the noise transmission of orthogonally stiffened panels were investigated with different numerical techniques [20, 21]. Statistical Energy Analysis was used to evaluate the transmission loss of a timber floor [22] and of a hybrid heavyweight-lightweight floor [23]. As the non resonant transmission is not described in the classical SEA model which describes energy-sharing between resonant modes, different authors have given particular attention to this aspect. Thus in [2] Crocker and Price described non resonant transmission through the panel by introducing a coupling loss factor between the excited and the receiving room. The parameter was estimated from the simple mass law equation of an infinite panel for frequencies below the critical frequency. In [24], the non resonant transmission through a double wall was studied. The sound was transmitted non-resonantly through each panel and it was pointed out that the cavity between the two panels cannot be considered as a semi-infinite free space. Indeed, this cavity is generally narrow compared to a wavelength and therefore only supports modes whose particle motion is parallel to the wall. Consequently, the conventional coupling loss factor considered by Crocker and Price [2] is poorly adapted, leading Craik [24] to propose a new formula. However, as the method proposed by these authors described non resonant transmission through an indirect coupling loss factor, the SEA model cannot predict the non resonant response of the structure. A methodology was presented in [25] by Renji et al. for estimating the non resonant response of the panel. It consisted in decomposing the panel into two SEA subsystems, one for the resonant contributions and one for the non resonant contributions. Although the effect of non resonant behaviour of the panel is predominant on sound transmission below the critical frequency, the authors found that the non resonant response of a limp panel (i.e. negligible panel stiffness) is small compared to the resonant response. On the contrary, for a thin light panel, the non resonant response can be significant at some frequencies.

In this paper, we propose to evaluate sound transmission through a complex panel separating two cavities by using the Statistical model Energy distribution Analysis (SmEdA) model [26-28]. This modal method is based on the same assumptions as the Statistical Energy Analysis (SEA) model except for the assumption of modal energy equipartition which is not assumed by the SmEdA model. It provides several advantages: (a) unlike classical SEA, it permits dealing with subsystems with low modal overlap driven by localized excitations [28] and a spatial distribution of the energy density inside each subsystem can be evaluated [29] for these cases; and (b) it uses the modal bases of uncoupled subsystems. These bases can be evaluated by using Finite Element models which permits dealing with subsystems having complex geometries. When modal energy equipartition is assumed in SmEdA, the SEA Coupling Loss Factors (CLF) can be deduced without any numerical matrix inversion. This parameter is estimated directly from an analytical expression depending on the modal information (i.e. natural frequency, modal damping, mode shape) contained in each uncoupled subsystem.
This approach was applied in [30] for estimating the coupling loss factors of a complex vibro-acoustic problem (i.e. train compartment). Like SEA, SmEdA describes in its original form the energy sharing by the resonant modes. Therefore only the resonant sound transmission through panels can be described. In this paper, we propose an extension of SmEdA for describing the contribution of the non resonant modes. This extension allows us to estimate the sound transmission of complex panels separating two non simply shaped cavities from modal information on the panel and the cavities. Hence, the influence of the cavity geometries and source location reported by different authors [4, 10-12] can be described as the effect of the mechanical and geometrical properties of the panel.

The present paper is organized as follows:
- The dual modal formulation (DMF) is used for describing the dynamic behaviour of a cavity-structure-cavity system in section 2. The resulting modal equations are the basis of the following developments. Using a numerical application in a test case, we highlight the influence of the low frequency non resonant modes of the structure on the sound transmission between the two cavities;
- The extension of SmEdA to the non resonant mode contributions is proposed in section 3. We obtain a new modal coupling scheme between the resonant modes of the three subsystems by condensing the amplitudes of the non resonant modes in the modal equations and by assuming that their behaviour is controlled by their mass. In this scheme, modal couplings controlled by stiffness elements appear between the cavity modes. The SmEdA formulation is then proposed based on this new modal coupling scheme.
- SmEdA including non resonant transmission is applied to a simple test case in section 4. For validation purposes, the noise reduction obtained with SmEdA is compared with that obtained by a direct resolution of the modal equation. The influence of damping is also studied
- Finally, an application for a cavity-ribbed plate-cavity system is presented to highlight the capacity of the approach to take into account the complexity encountered in industrial applications.

## 2. Dual modal formulation (DMF)

DMF can be used for calculating the response of coupled subsystems from prior knowledge of the modes of each uncoupled subsystem. This formulation, which has been well-known for decades, is suitable for describing the dynamic behaviour of a flexible panel coupled with acoustic cavities. A Green formulation [1] or a variational formulation of the fluid-structure problem can be used to obtain the modal equation of motion. DMF has also been extended to the general case of the coupling of two elastic continuous mechanic systems [26]. The modal interaction scheme obtained with this formulation is in accordance with the mode coupling assumed in classical SEA and is also the basis of SmEdA. After having described the cavity-structure-cavity system considered in this paper, we present the DMF results and discuss the convergence of the modal series on the basis of a numerical test case.

### *2.1 Cavity-structure-cavity system*

Let us consider the cavity-structure-cavity problem shown in Fig. 1. It is composed of two air cavities of volumes $V_1$ and $V_2$, and an elastic thin structure of surface $S$. The two cavities can exchange vibrating energy through the thin structure, whereas all the other cavity walls are assumed to be rigid. Thus the fluid-structure interface corresponds to surface $S$. The thickness, mass density and damping loss factor of the thin structure are denoted $h$, $\rho$, $\eta_S$ respectively, and $c_0$, $\rho_0$ are the air celerity and mass density. $\eta_{C1}$, and $\eta_{C2}$ represent the damping loss factor of cavities 1 and 2 respectively.

To estimate the noise reduction (NR) of the panel, a single acoustic source placed in cavity 1 is considered. To simplify our presentation, we assume that this acoustic source is a monopole located at point $M_0$ with strength $Q_0$ (i.e. volume velocity [1]). The auto-spectrum density of the source strength, $S_{Q_0 Q_0}$ is assumed constant (i.e. white spectrum) in the frequency band $[\omega_1, \omega_2]$ of central frequency $\omega_c$.

As the method developed in this paper provides the energy, the noise transmission will be characterised by the Energy Noise Reduction (ENR) defined by:

$$ENR = 10 \log_{10} \left( \frac{\langle E_{C1} \rangle_t}{\langle E_{C2} \rangle_t} \right), \qquad (1)$$

where $\langle E_{C1} \rangle_t$ and $\langle E_{C2} \rangle_t$ are the time-averaged total energies (i.e. kinetic energy + strain energy) of cavity 1 and cavity 2, respectively.

This parameter is related to the classical Transmission Loss (TL) by:

$$TL = ENR - 10 \log \left( \frac{V_1}{V_2} \right) - 10 \log \left( \frac{4 \eta_{C2} \omega_c V_2}{c_0 S} \right). \qquad (2)$$

In the next section, we present the main results of the dual modal formulation applied to the present case.

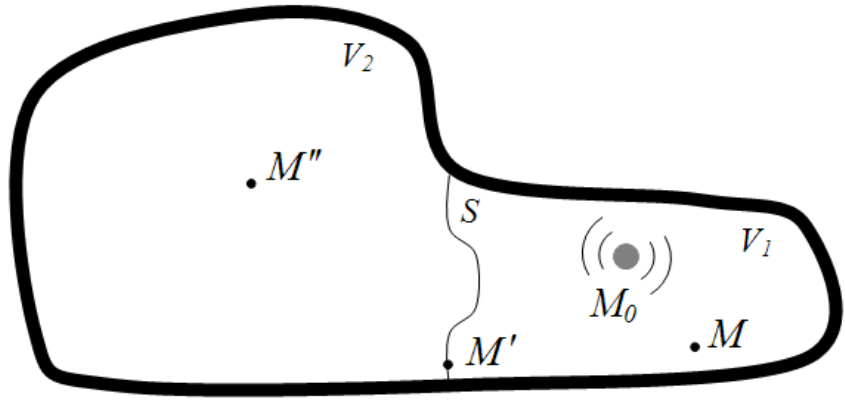

Figure 1. Illustration of the sound transmission through a thin light structure.

### 2.2 DMF equations

In accordance with the DMF, the panel is described by its displacement field (i.e. normal displacement) and uncoupled-free modes (i.e. in-vacuo modes of the structure) whereas the cavities are described by stress fields (i.e. acoustic pressure) and by their uncoupled-blocked modes (i.e. rigid wall modes of the cavity). The boundary conditions of these uncoupled subsystem modes are illustrated in Fig. 2 for the present case. These subsystem modes can be easily calculated analytically for academic cases [8, 31], or numerically with Finite Element models for complex cases [30].

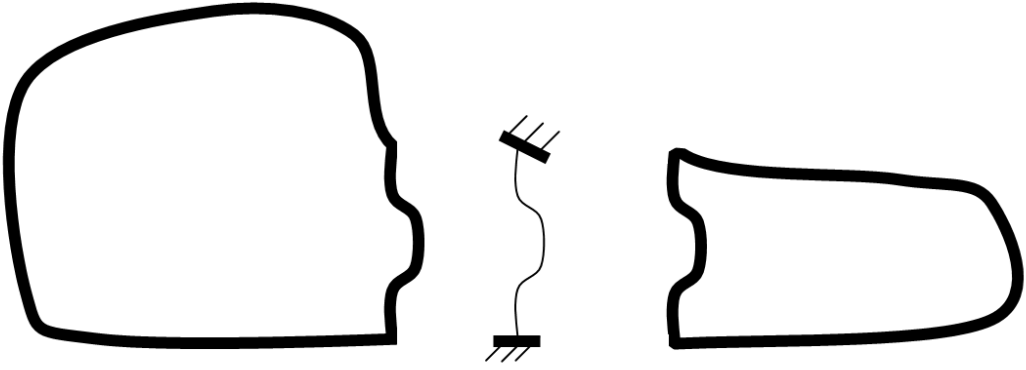

Figure 2. DMF substructuring.

The modal expansions of the normal displacement $W$ at point $M'$ on the panel may be written

$$W(M',t) = \sum_{q=1}^{\infty} \chi_q(t) \tilde{W}_q(M'), \qquad (3)$$

whereas the acoustic pressures $p$ at points $M$ and $M''$ inside cavities 1 and 2, respectively, are written as:

$$p(M,t) = \sum_{p=1}^{\infty} \phi_p(t) \tilde{p}_p(M), \qquad (4)$$

$$p(M'',t) = \sum_{r=1}^{\infty} \xi_r(t) \tilde{p}_r(M''), \qquad (5)$$

where:

- $\phi_p$, $\chi_q$, $\xi_r$ are modal amplitudes ;

- $\tilde{p}_p$, $\tilde{p}_r$ are the pressure mode shapes of cavities 1 and 2, respectively. For the sake of convenience, these mode shapes are normalised to a unit modal stiffness (i.e. $\frac{1}{\rho_0 c_0^2} \int_{\Omega_1} \tilde{p}_p^2 dV = 1$, $\frac{1}{\rho_0 c_0^2} \int_{\Omega_1} \tilde{p}_r^2 dV = 1$);

- $\tilde{W}_q$ are the displacement mode shapes of the panel, normalised to a unit modal mass (i.e. $\int_S \rho h \tilde{W}_q^2 dS = 1$).

Thereafter, the space and time dependencies are deleted from the notations, although they are still considered. DMF consists in introducing these expansions (3-5) in a weak formulation of the vibro-acoustic problem considered and using the orthogonality properties of the uncoupled modes. For more details on this formulation, the reader can refer to [1, 26].

Finally, with the change of modal variables (where the prime symbol indicates the time derivative),

$$\phi_p = \varphi'_p, \ \forall p \in \square^*, \ \xi_r = \zeta'_r, \ \forall r \in \square^* \qquad (6)$$

The modal equations can be written in the following form,

$$\begin{cases} \varphi_p'' + \omega_p \eta_p \varphi_p' + \omega_p^2 \varphi_p - \sum_{q=1}^{\infty} W_{pq} \chi_q' = Q_p, \quad \forall p \in \mathbb{N}^* \\ \chi_q'' + \omega_q \eta_q \chi_q' + \omega_q^2 \chi_q + \sum_{p=1}^{\infty} W_{pq} \varphi_p' + \sum_{r=1}^{\infty} W_{rq} \zeta_r' = 0, \quad \forall q \in \mathbb{N}^* \\ \zeta_r'' + \omega_r \eta_r \zeta_r' + \omega_r^2 \zeta_r - \sum_{q=1}^{\infty} W_{rq} \chi_q' = 0, \quad \forall r \in \mathbb{N}^* \end{cases} \quad (7)$$

where: - $Q_p$ are the generalised source strengths due to the acoustic source $Q_0$;

- $\omega_p$, $\omega_q$, $\omega_r$ are the angular frequencies of subsystem modes;

- $\eta_p$, $\eta_q$, $\eta_r$ are the modal damping loss factors (for which we assume $\eta_p = \eta_{C1}$, $\eta_q = \eta_S$, $\eta_r = \eta_{C2}$), and

- $W_{pq}, W_{qr}$ are the modal interaction works defined by:

$$W_{pq} = \int_S \tilde{W}_q \tilde{p}_p dS, \; W_{qr} = \int_S \tilde{W}_q \tilde{p}_r dS. \quad (8)$$

The form of these equations allows us to interpret mode interactions as oscillators coupled by gyroscopic elements (introducing coupling forces proportional to the oscillators' velocities and of opposite signs, [1]). Note that a mode of one subsystem is coupled to the modes of the other subsystem but is not directly coupled with the other modes of the subsystem to which it belongs.

In theory, the modal summations of these equations have an infinite number of terms. In practice, only a finite number of modes can be considered. In section 2.2 we discuss the choice of these modes to obtain a reliable estimation of the response of the cavity-structure-cavity system. In the meantime, let us consider a finite set of modes for each subsystem. We note the mode sets as $\hat{P}$, $\hat{Q}$, and $\hat{R}$ for cavity 1, the structure and cavity 2, respectively. $P$, $Q$, $R$ represent the number of modes contained in these sets.

Let us apply the Fourier transform to (7). By considering the finite sets of modes, we obtain the following matrix system:

$$\begin{bmatrix} Z_{11} & -j\omega W_{12} & 0 \\ +j\omega W_{12}^* & Z_{22} & +j\omega W_{23}^* \\ 0 & -j\omega W_{23} & Z_{33} \end{bmatrix} \begin{bmatrix} \Gamma_1 \\ \Gamma_2 \\ \Gamma_3 \end{bmatrix} = \begin{bmatrix} Q_1 \\ 0 \\ 0 \end{bmatrix} \quad (9)$$

with the modal amplitude and generalised source strength vectors:

$$\Gamma_1 = \begin{bmatrix} \vdots \\ \bar{\varphi}_p \\ \vdots \end{bmatrix}_{P \times 1}, \; \Gamma_2 = \begin{bmatrix} \vdots \\ \bar{\chi}_q \\ \vdots \end{bmatrix}_{Q \times 1}, \; \Gamma_3 = \begin{bmatrix} \vdots \\ \bar{\zeta}_r \\ \vdots \end{bmatrix}_{R \times 1}, \; Q_1 = \begin{bmatrix} \vdots \\ \bar{Q}_p \\ \vdots \end{bmatrix}_{P \times 1}, \quad (10)$$

the modal impedance matrices,

$$Z_{11} = diag\left[-\omega^2 + j\omega\omega_p \eta_p + \omega_p^2\right]_{P \times P}, \; Z_{22} = diag\left[-\omega^2 + j\omega\omega_q \eta_q + \omega_q^2\right]_{Q \times Q} \quad (11)$$

$$Z_{33} = diag\left[-\omega^2 + j\omega\omega_r \eta_r + \omega_r^2\right]_{R \times R}, \text{ and,} \quad (12)$$

the modal interaction work matrices,

$$W_{12} = [W_{pq}]_{P \times Q}, \; W_{23} = [W_{qr}]_{Q \times R}. \quad (13)$$

The asterisk in (9) indicates the transpose of the (real) matrix.

## 2.3 Resolution of the DMF equations

The DMF equations (9-13) will be used in section III to develop the SmEdA model including the non resonant transmission. These equations can also be solved directly to calculate the pure tone total energy of each cavity. Although these DMF calculations are more time-consuming than the SmEdA calculation, they give a point of comparison by summing up the energies at frequencies in the band of excitation. Here we present the outline of this resolution.

It is assumed that the acoustic source presents a white noise spectrum in the frequency band $[\omega_1, \omega_2]$. Then, the time average of the total energy of cavity $i \in \{1,2\}$, $<E_{Ci}>_t$ can be obtained by

$$<E_{Ci}>_t = S_{Q_0 Q_0} \int_{\omega_1}^{\omega_2} E_{Ci}(\omega) \quad , \tag{14}$$

where $E_{Ci}(\omega)$ is the harmonic total energy of cavity $i$ when the system is excited by the monopole at $M_0$ with a unit strength amplitude (i.e. unit volume velocity). It can be estimated by solving the DMF equations with a generalised source strength vector given by:

$$Q_1 = \begin{bmatrix} \vdots \\ -\rho_0 c_0^2 \tilde{p}_p(M_0) \\ \vdots \end{bmatrix}_{P \times 1} \tag{15}$$

As the number of cavity modes (i.e. $P$ and $R$) rapidly becomes much higher than the number of plate modes when the frequency increases, it is advisable to condense the modal amplitude vectors related to the cavity modes $\Gamma_1$ and $\Gamma_2$ in the matrix system (9). The modal amplitude vector related to the structure modes, $\Gamma_2$ can be calculated from:

$$\Gamma_2 = -j\omega Z_{22}'^{-1} W_{12}^* Z_{11}^{-1} Q_1 \quad , \tag{16}$$

with $Z_{22}' = Z_{22} - \omega^2 W_{12}^* Z_{11}^{-1} W_{12} - \omega^2 W_{23}^* Z_{33}^{-1} W_{23}$. \hfill (17)

As the modal impedance matrices, $Z_{11}$ and $Z_{22}$ are diagonal, their inverses - which appear in the previous equation - can be obtained immediately. Eq. (16) then requires the numerical inversion of a square matrix whose dimension is the number of plate modes (i.e. $Q$).

The modal amplitude vectors related to the cavity modes are then deduced with:

$$\Gamma_1 = Z_{11}^{-1}\left(j\omega W_{12}\Gamma_2 + Q_1\right), \quad \Gamma_3 = j\omega Z_{33}^{-1} W_{23}\Gamma_2. \tag{18}$$

The total energy of each mode can be deduced from its modal amplitude (see [32]). The total energy of each subsystem is then obtained by summing the energies of its subsystem modes (taking the property of orthogonality of the subsystem modes into account). For more details on this aspect, the reader can refer to [26].

The process described previously allows us to estimate $E_{Ci}(\omega)$. The time average of the total energy of cavity $i \in \{1,2\}$, $<E_{Ci}>_t$ is obtained from a numerical estimation of the integral of (14) using the rectangular rule with a frequency step defined in accordance with the smallest damping bandwidths of the different subsystems.

## *2.4. Convergence of modal expansions*

In this section, we study a test case to evaluate the accuracy of the DMF results as a function of the modes considered in the modal expansions. The test case considered is the cavity-plate-cavity system illustrated in Fig. 3. It is composed of a rectangular simply-supported plate coupled on both sides with a parallelepiped cavity. The dimensions of the steel plate are 0.8 m x 0.6m, with a thickness of 1mm (mass density $\rho=7800$ kg/m$^3$, Young modulus $E=2.10^{11}$Pa, $\eta_S=0.01$).The cavity is filled with air (mass density $\rho_0=1.29$ kg/m$^3$, speed of sound $c_0=340$ m/s, damping loss factor $\eta_{C1}=\eta_{C2}=0.01$). The length and width of the two cavities are the same as those of the plate. Cavity 1 has a depth of 0.8m whereas cavity 2 has a depth of 0.7m. The behaviour of the plate will be described by the Kirchhoff equation (thin plate) whereas the Helmholtz equation will be considered for the acoustic domain. Cavity 1 is assumed to be excited by a harmonic monopole source at (0.24 m, 0.42 m, 0.54 m) in the coordinate system *(O,x,y,z)* represented in Fig. 3. The source strength is set to a unit volume velocity.

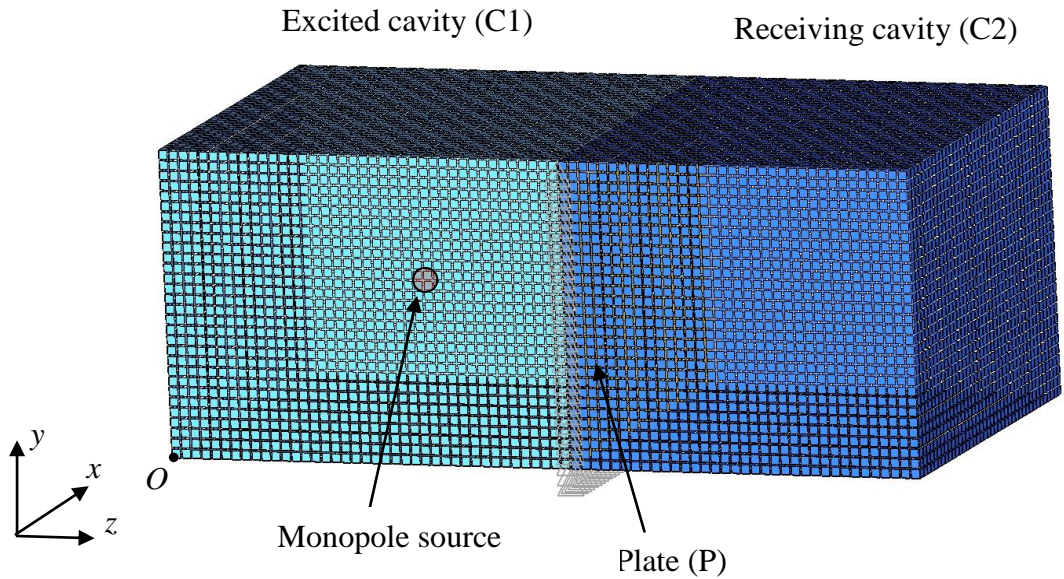

Figure 3. Finite Element model of the test case (99138 Nodes, 1200 CQUAD4 2D elements, 90000 CHEXA 3D elements).

A reference result for the present test case is obtained by using the Finite Element (FE) method and the MD NASTRAN code. The FE mesh shown in Fig. 3 has been defined to allow calculation up to 750 Hz (by considering a criterion of 6 elements per flexural wavelength). The harmonic response is obtained by direct analysis (SOL108 in NASTRAN code) and the subsystem energies are estimated from the node responses using a homemade MATLAB code.
For the present test case, the subsystem modal information required in the DMF equations (16-18) can be calculated analytically for both the natural frequencies and the interaction modal works (see [31]). Two DMF calculations are performed to estimate the response in the 500 Hz octave band: only the resonant modes of the three subsystems are considered in the first while the second also considers the low frequency non resonant modes of the plate. Comparisons of these two DMF results with the FEM reference result are proposed in Fig. 4.

Good agreement can be seen between the three calculations of the energy of the excited cavity, whereas the energy of the receiving cavity is not correctly predicted by the DMF calculation, which considers only the resonant modes. This result is well-known and shows that the resonant mode transmission considered in the SEA and SmEdA models is unable to describe the behaviour correctly in the case considered. Conversely, the result of the DMF calculation of the resonant modes and the low frequency non resonant plate modes is very close to the reference result, even for the receiving cavity. The Non Resonant (NR) modes considered in the second DMF calculation play an important role in sound transmission for frequencies below the critical frequency of the plate, which in this case is around 11.7 kHz. The importance of these NR modes can be explained in the DMF equations through the modal interaction works. Fig. 5 shows a plot of the interaction works between the Resonant cavity modes and Resonant/Non Resonant plate modes for the 500 Hz octave band. The highest values are observed for the NR plate modes with a low modal order, corresponding to modal frequencies below 100 Hz. These modes are not in frequency coincidence with the excited resonant cavity modes but they are in shape coincidence with these cavity modes. This result indicates also that the NR cavity modes and the high frequency NR plate modes can be neglected as they are not in the coincidence frequency and are spatially poorly coupled. Globally, for the octave band 500 Hz, the ENR (i.e. Eq. (1)) predicted by the FEM is 21.1 dB whereas the DMF gives 20.1 dB and 28.8 dB, with and without the NR plate modes, respectively. The 1dB difference between the FEM and DMF calculations with the NR modes is certainly due to the slight difference of modelling (FE discretisation, modelling of the plate using shell elements). This result is satisfactory, however, and validates the DMF calculation. In the following part of the paper, the DMF calculation with the NR plate modes will be considered as the reference.

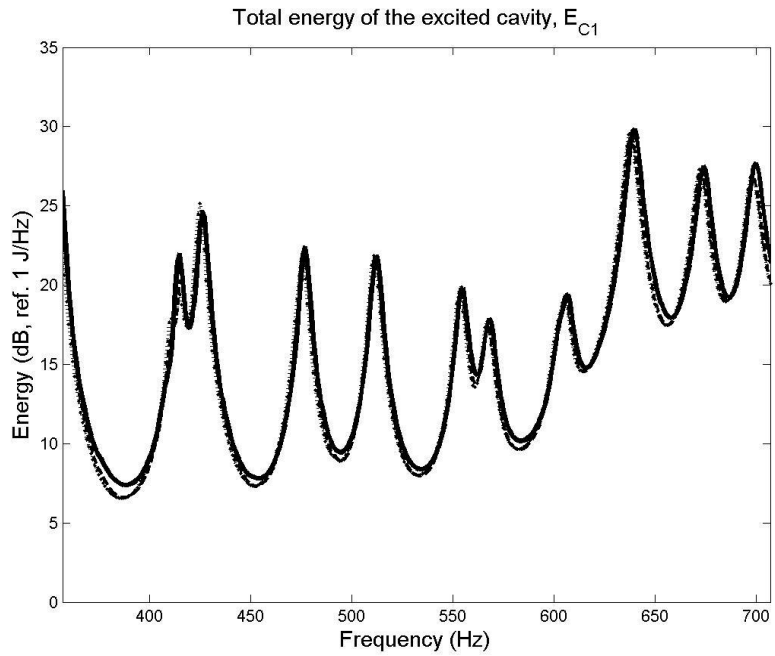

(a)

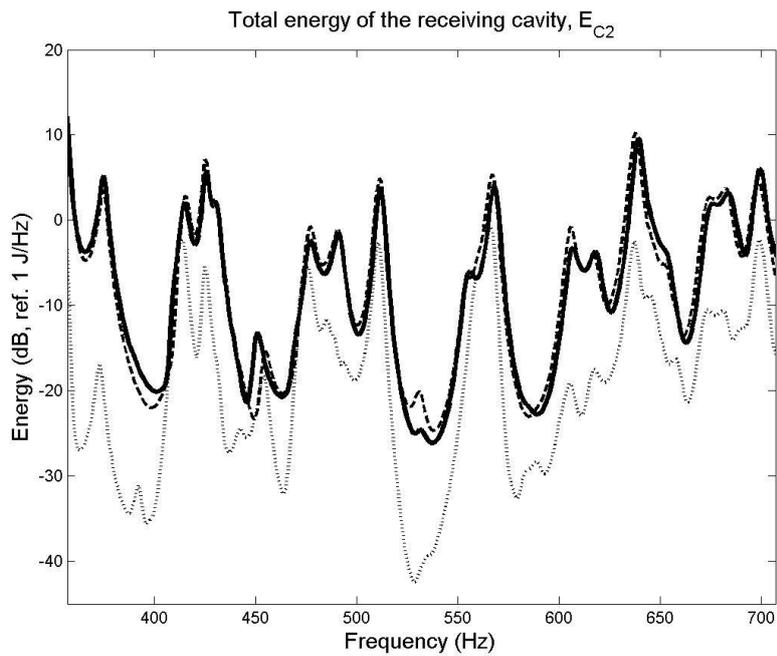

(b)

Figure 4. Total energy in the octave band 500 Hz for : (a), the excited cavity; (b), the receiving cavity. Comparison of three calculations: solid line, FEM (reference); dashed line, DMF with NR plate mode; dotted line, DMF without NR plate mode;

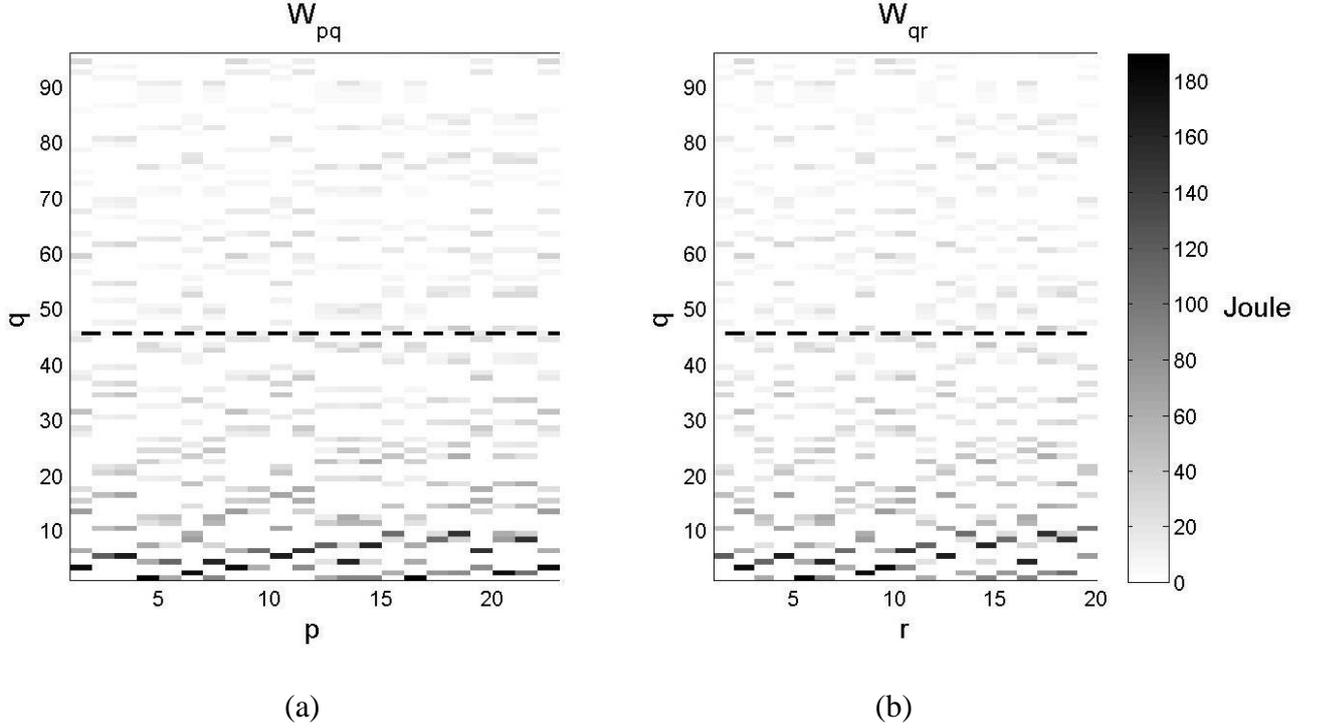

Figure 5. Interaction of modal works (in Joules); (a), between the excited cavity and the plate (i.e. $W_{pq}$); (b), between the plate and the receiving cavity (i.e. $W_{qr}$). Results for the octave band 500 Hz. Dashed line bound between the resonant and non resonant plate modes.

In order to study the influence of the NR plate modes on energy noise reduction in the frequency range [400 Hz – 20 kHz], DMF calculations with and without NR plate modes were performed for each third octave band in this frequency domain. For each band, Tab. 1 summarises the number of Resonant modes and Non Resonant modes taken into account in the DMF calculations. It can be seen that the number of Resonant cavity modes is small (less than 10) for the first 1/3 octave bands. For the last 1/3 octave bands, the numbers of Resonant cavity modes are prohibitive but the DMF calculations were performed despite that fact that the total number of modes was greater than 400 000 as the resolution proposed with Eq. (16-18) only requires the numerical inversion of a square matrix whose dimension is the number of plate modes.

The ENR results are proposed in Fig. 6 for each third octave band. The results of the previous FEM calculation for the three first third octave bands were also plotted (by crosses) for information.

The discrepancies between the two DMF calculations are greater than 8 dB below 10 kHz, indicating the predominance of NR transmission for these frequencies. Then, the ENR falls for the 12.5 kHz third octave band due to the coincidence phenomenon (the critical frequency is around 11.7 kHz). For frequencies above 10 kHz, the differences between the two DMF results are negligible, indicating that the NR plate modes play an insignificant role. These tendencies agree with previous studies on the noise transmission of panels, demonstrating that DMF is a fair representation of the system considered and proves to be a good tool for investigating the sound transmission at low frequency if the NR plate modes are taken into account. An illustration of the resulting modal coupling scheme is given in Fig. 7. In the next section we consider it as the base of the SmEdA formulation integrating the non resonant transmission.

|  | 400 Hz | 500 Hz | 630 Hz | 800 Hz | 1 kHz | 1.25kHz | 1.6 kHz | 2 kHz | 2.5 kHz |
|---|---|---|---|---|---|---|---|---|---|
| $P$ | 5 | 6 | 12 | 22 | 41 | 71 | 149 | 263 | 535 |
| $Q^{NR}$ | 46 | 59 | 75 | 96 | 124 | 157 | 198 | 251 | 322 |
| $Q^{R}$ | 13 | 16 | 21 | 28 | 33 | 42 | 52 | 70 | 83 |
| $R$ | 4 | 4 | 12 | 21 | 32 | 67 | 129 | 231 | 472 |

|  | 3.15kHz | 4 kHz | 5 kHz | 6.3 kHz | 8kHz | 10kHz | 12.5kHz | 16kHz | 20kHz |
|---|---|---|---|---|---|---|---|---|---|
| $P$ | 1033 | 1998 | 3982 | 7815 | 15490 | 30672 | 60818 | 121228 | 236518 |
| $Q^{NR}$ | 406 | 515 | 655 | 829 | 1049 | 1331 | 1682 | 2122 | 2687 |
| $Q^{R}$ | 108 | 139 | 173 | 219 | 281 | 350 | 439 | 564 | 703 |
| $R$ | 909 | 1766 | 3483 | 6859 | 13560 | 26876 | 53361 | 106085 | 209151 |

Tab. 1. Numbers of modes for each third octave band: $P$, number of modes for the excited cavity; $Q^{NR}$, number of non-resonant modes of the plate; $Q^{R}$, number of resonant modes of the plate; $R$, number of modes for the receiving cavity.

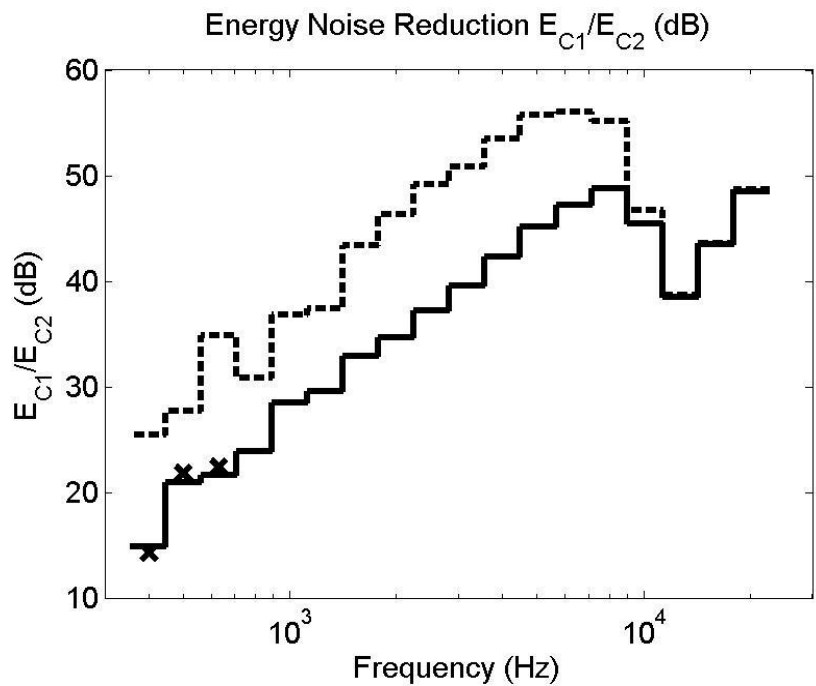

Figure 6. Energy Noise Reduction $E_{C1}/E_{C2}$ versus third octave band. Comparison of three calculations: cross, FEM; solid line, DMF with NR plate modes and border modes; dashed line, DMF without NR plate modes.

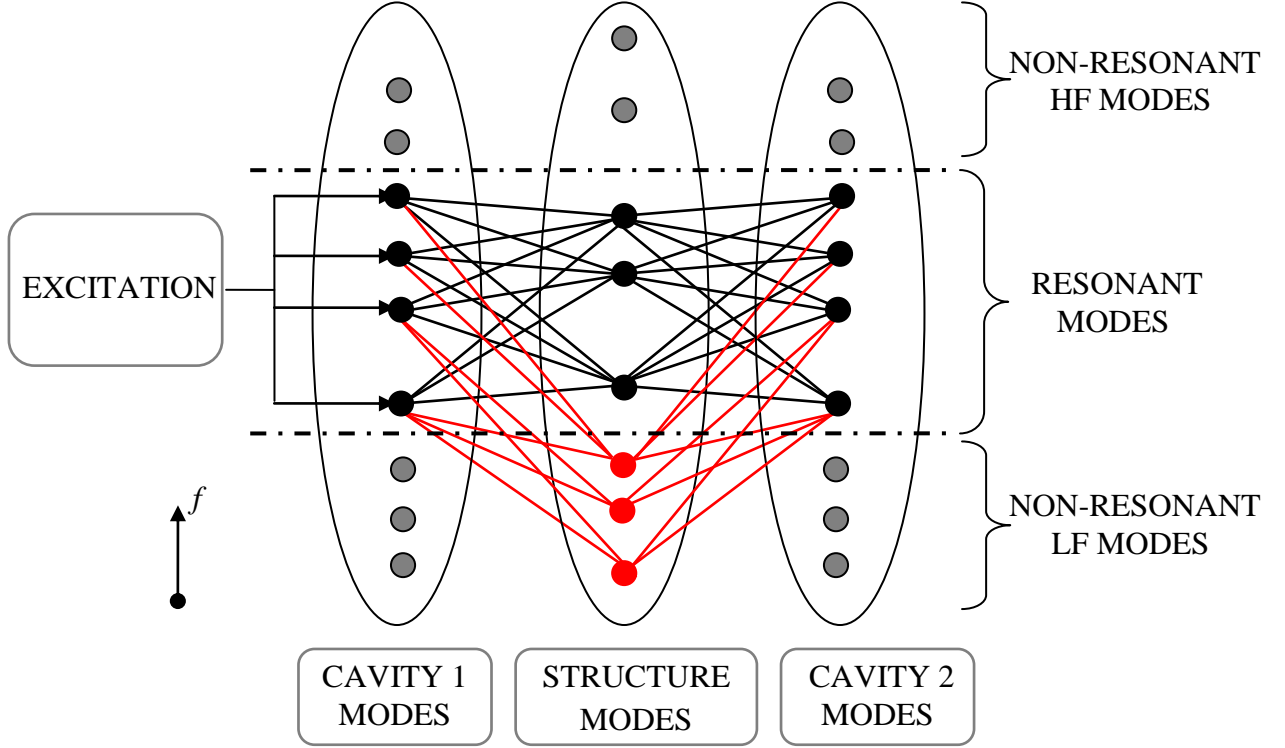

Figure 7. Modal interaction scheme of the sound transmission through a thin light structure, below the critical frequency.

## 3. SmEdA model with non resonant mode transmission

This section consists of the energy formulation of the vibro-acoustic behaviour of the cavity-structure-cavity system when the acoustic source is characterised by a white noise spectrum in the frequency band $[\omega_1, \omega_2]$. It is organised as follows: first, manipulations of the DMF equations and approximations of terms are considered, leading to a new modal interaction scheme; then, the power flow relation involving two coupled oscillators is recalled and, finally, the energy equations of motions of the problem considered are written on the basis of the new modal interaction scheme and the oscillator power flow relation.

In the following developments, the brackets indicating the time-averaged energy will be omitted from the notations although this average is always considered.

### *3.1 Non-resonant mode condensation*

A new interpretation of the modal coupling shown in Fig. 7 is proposed in this section. For that, let us consider two sets of modes for the structure: the Resonant modes set, $\hat{Q}^R$ and the Non Resonant modes set, $\hat{Q}^{NR}$ which are defined by:

$$\begin{cases} q \in \hat{Q}^{NR} \Leftrightarrow \omega_q \in [0, \omega_1[ \\ q \in \hat{Q}^R \Leftrightarrow \omega_q \in [\omega_1, \omega_2] \end{cases} \quad (19)$$

The modal impedance and interaction work matrices can be written as

$$Z_{22} = \begin{bmatrix} Z_{22}^{NR} & 0 \\ 0 & Z_{22}^{R} \end{bmatrix}, \quad W_{12} = \begin{bmatrix} W_{12}^{NR} & W_{12}^{R} \end{bmatrix}, \quad W_{23} = \begin{bmatrix} W_{23}^{NR} & W_{23}^{R} \end{bmatrix}, \quad (20)$$

where subscript R or NR associated with the sub-matrices are related to the resonant and non-resonant mode sets, respectively.

The DMF equations in the frequency domain (9) can be rewritten:

$$\begin{bmatrix} Z_{11} & -j\omega W_{12}^{NR} & -j\omega W_{12}^{R} & 0 \\ +j\omega W_{12}^{NR*} & Z_{22}^{NR} & 0 & +j\omega W_{23}^{NR*} \\ +j\omega W_{12}^{R*} & 0 & Z_{22}^{R} & +j\omega W_{23}^{R*} \\ 0 & -j\omega W_{23}^{NR} & -j\omega W_{23}^{R} & Z_{33} \end{bmatrix} \begin{bmatrix} \Gamma_1 \\ \Gamma_2^{NR} \\ \Gamma_2^{R} \\ \Gamma_3 \end{bmatrix} = \begin{bmatrix} Q_1 \\ 0 \\ 0 \\ 0 \end{bmatrix} \quad (21)$$

Eliminating $\Gamma_2^{NR}$ from the second row, the following equations are derived,

$$\Gamma_2^{NR} = -j\omega \left[ Z_{22}^{NR} \right]^{-1} \left( W_{12}^{NR*} \Gamma_1 + W_{23}^{NR*} \Gamma_3 \right), \quad (22)$$

which can be reintroduced in the other rows to give the condensed matrix system:

$$\begin{bmatrix} Z_{11} - \omega^2 W_{12}^{NR} \left[ Z_{22}^{NR} \right]^{-1} W_{12}^{NR*} & -j\omega W_{12}^{R} & -\omega^2 W_{12}^{NR} \left[ Z_{22}^{NR} \right]^{-1} W_{23}^{NR*} \\ +j\omega W_{12}^{R*} & Z_{22} & +j\omega W_{23}^{R} \\ -\omega^2 W_{23}^{NR} \left[ Z_{22}^{NR} \right]^{-1} W_{12}^{NR*} & -j\omega W_{23}^{R} & Z_{33} - \omega^2 W_{23}^{NR} \left[ Z_{22}^{NR} \right]^{-1} W_{23}^{NR*} \end{bmatrix} \begin{bmatrix} \Gamma_1 \\ \Gamma_2^{R} \\ \Gamma_3 \end{bmatrix} = \begin{bmatrix} Q_1 \\ 0 \\ 0 \end{bmatrix} \quad (23)$$

This operation allows us to link together the amplitudes of the resonant modes of the 3 subsystems. On the other hand, if we consider that the natural frequencies of the NR modes are much lower than the angular frequency (i.e. $\omega_q \ll \omega_1 < \omega, \forall q \in \hat{Q}^{NR}$), it can be assumed that:

$$\left[ Z_{22}^{NR} \right]^{-1} \approx -\frac{1}{\omega^2} I, \quad (24)$$

where $I$ is the identity matrix $Q^{NR} \times Q^{NR}$. It should not be forgotten here that the modal masses were normalized to unity.

Moreover, if weak coupling is assumed between the plate and a cavity, it is possible to neglect the added terms modifying the modal impedance matrices of the cavities (i.e. terms after $Z_{11}$ and $Z_{33}$ in Eq. (23)), we obtain:

$$\begin{bmatrix} Z_{11} & -j\omega W_{12}^{R} & -W_{12}^{NR} W_{23}^{NR*} \\ +j\omega W_{12}^{R*} & Z_{22} & +j\omega W_{23}^{R*} \\ -W_{23}^{NR} W_{12}^{NR*} & -j\omega W_{23}^{R} & Z_{33} \end{bmatrix} \begin{bmatrix} \Gamma_1 \\ \Gamma_2^{r} \\ \Gamma_3 \end{bmatrix} = \begin{bmatrix} Q_1 \\ 0 \\ 0 \end{bmatrix} \quad (25)$$

As will be shown in the numerical applications of section IV, the effects of these added terms is negligible when the cavities are filled with air. This would not be the case if the cavities were filled with water (as illustrated in [32]).

Applying the inverse Fourier transform to (25) allows us to write these equations in the time domain:

$$\begin{cases} \varphi_p'' + \omega_p \eta_p \varphi_p' + \omega_p^2 \varphi_p - \sum_{q \in \hat{Q}^R} W_{pq} \chi_q' - \sum_{r \in \hat{R}} \left( \sum_{q \in \hat{Q}^{NR}} W_{pq} W_{rq} \right) \chi_r = Q_p, \ \forall p \in \hat{P} \\ \chi_q'' + \omega_q \eta_q \chi_q' + \omega_q^2 \chi_q + \sum_{p \in \hat{P}} W_{pq} \varphi_p' + \sum_{r \in \hat{R}} W_{rq} \zeta_r' = 0, \ \forall q \in \hat{Q}^R \\ \zeta_r'' + \omega_r \eta_r \zeta_r' + \omega_r^2 \zeta_r - \sum_{q \in \hat{Q}^R} W_{rq} \chi_q' - \sum_{p \in \hat{P}} \left( \sum_{q \in \hat{Q}^{NR}} W_{pq} W_{rq} \right) \varphi_p = 0, \ \forall r \in \hat{R} \end{cases} \quad (26)$$

This system is used to give a new interpretation of the modal interaction compared to the scheme proposed in Fig. 7. The resonant modes of the structure remain connected to the resonant modes of each cavity by gyroscopic elements. The non resonant modes are no longer represented explicitly, but their effects are represented through direct couplings between the resonant modes of the two cavities. The coupling elements introduce a coupling force proportional to the modal amplitudes (and not to their time derivatives) which is typical of a stiffness effect. The cavity modes are then coupled by springs whose stiffness is given by the modal interaction between the resonant cavity modes and the non resonant structure modes (i.e. $\sum_{q \in \hat{Q}^{NR}} W_{pq} W_{rq}$). Surprisingly, the mass control behaviour of the non resonant structure modes leads to stiffness couplings with the cavity modes. This is due to the dual displacement-stress formulation of the problem: the cavities are described by their pressure field (i.e. their stresses) whereas the structure is described by its displacements ([26]). The interpretation in the equations of motion of a spring or a mass element (from the time derivative) may change as a function of the descriptive variables (i.e. stress/displacement) considered in the equations.

An illustration of the new modal interaction scheme is proposed in Fig. 8. Compared to the scheme proposed in Fig. 7, it has the advantage of involving only couplings between resonant modes. The same process as that used to establish the original SmEdA model can now be applied to the present problem. The energy formulation can be based on the power flow relation for two coupled oscillators excited by a white noise source. This relation and its assumptions are recalled in the next section.

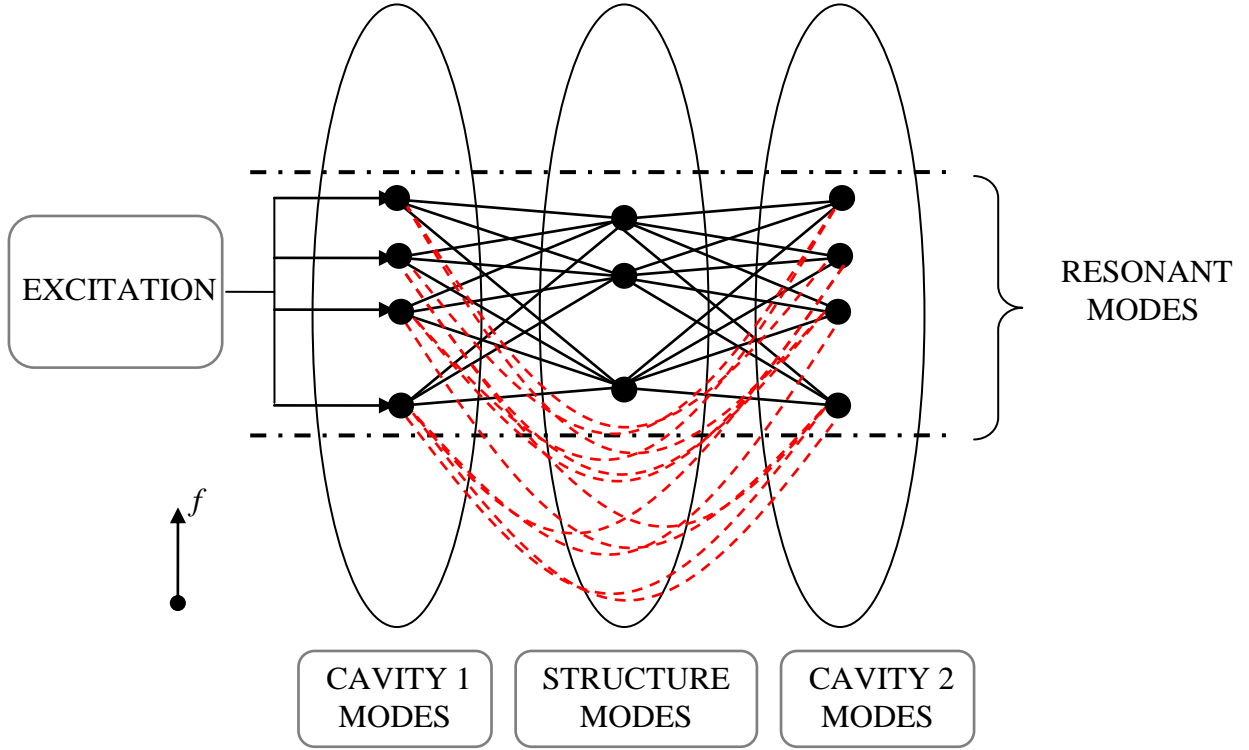

Figure 8. Interpretation of the resonant modal interaction for the sound transmission through a thin light structure below the critical frequency. Solid line, gyroscopic coupling; dashed line, spring coupling.

### *3.2 Energy sharing between two coupled oscillators*

As shown in Fig. 9, we consider two oscillators coupled via a spring and a gyroscopic element. These coupling elements are characterised by their constants of proportionality $\kappa$ and $\gamma$ for the spring and the gyroscopic element, respectively. Without loss of generality, we consider a unit mass for each oscillator. Once again we recall that the modal masses were normalized to unity. $K_1$, $K_2$ are the stiffnesses of the oscillators. The natural angular frequencies of each uncoupled-blocked oscillator are therefore $\bar{\omega}_1 = \sqrt{K_1 + \kappa}$ and $\bar{\omega}_2 = \sqrt{K_2 + \kappa}$, for oscillators 1 and 2, respectively.

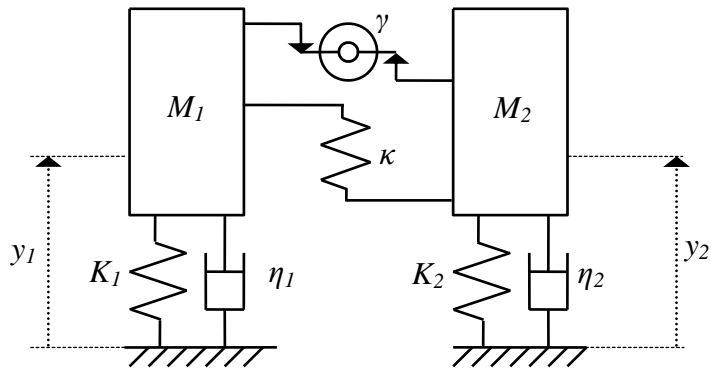

Figure 9. Two oscillators coupled by spring and gyroscopic elements.

Each oscillator is damped by a viscous absorber of damping loss factors: $\eta_1$ for oscillator 1 and $\eta_2$ for oscillator 2. The equations of motion for the two coupled oscillators excited by external forces $F_1$ and $F_2$ are then yielded by:

$$\begin{cases} y_1'' + \bar{\omega}_1 \eta_1 y_1' + \bar{\omega}_1^2 y_1 - \gamma y_2' - \kappa y_2 = F_1, \\ y_2'' + \bar{\omega}_2 \eta_2 y_2' + \bar{\omega}_2^2 y_2 + \gamma y_1' - \kappa y_1 = F_2. \end{cases} \quad (27)$$

Moreover, we assume that the external forces are independent (uncorrelated), stationary and have a constant power spectrum density for all frequencies (white noise). It has been demonstrated in this case [33] that the time-averaged power flow from oscillator 1 to oscillator 2, $\Pi_{12}$, is proportional to the difference of the time-averaged total energies of the oscillators $(E_1 - E_2)$ that is:

$$\Pi_{12} = \beta(E_1 - E_2), \quad (28)$$

where coefficient $\beta$ is expressed by:

$$\beta = \frac{\gamma^2 \left(\bar{\omega}_1 \eta_1 \bar{\omega}_2^2 + \bar{\omega}_2 \eta_2 \bar{\omega}_1^2\right) + \kappa^2 \left(\bar{\omega}_1 \eta_1 + \bar{\omega}_2 \eta_2\right)}{\left(\bar{\omega}_1^2 - \bar{\omega}_2^2\right)^2 + \left(\bar{\omega}_1 \eta_1 + \bar{\omega}_2 \eta_2\right)\left(\bar{\omega}_1 \eta_1 \bar{\omega}_2^2 + \bar{\omega}_2 \eta_2 \bar{\omega}_1^2\right)} \quad (29)$$

It can be seen that $\beta$ depends on the natural angular frequencies of the uncoupled oscillators, the damping loss factors and the coupling coefficients, $\kappa$ and $\gamma$.

This expression is valid for uncorrelated external excitations with white noise spectrums. Nevertheless, it also remains a fair approximation if the external excitations have white noise spectrums in a frequency band containing the natural frequencies (and a null spectrum outside this band) ([33-35]). Hence, in the next section, we could use it to estimate the power flow exchanged by two resonant modes for external excitations in the frequency band $[\omega_1, \omega_2]$.

We emphasize that relation (28) supposes white spectrum forces in the frequency band $]-\infty, +\infty[$. It remains a reliable approximation for white spectrum forces in a band $[\omega_1, \omega_2]$, if the blocked natural angular frequencies of the two oscillators $\bar{\omega}_1$, $\bar{\omega}_2$ are contained in this band (i.e. if the two oscillators are resonant). Indeed, in this case, the spectra (i.e. frequency decomposition) of the oscillator energies or the power flow exhibit significant values only for frequencies "close to" the blocked natural frequencies of the oscillators. Thus, the angular frequencies outside the band $[\omega_1, \omega_2]$ have a negligible contribution on the evaluation of the time-averaged energies and the time-averaged power flow even if the forces have a white spectrum for frequency from $-\infty$ to $+\infty$ (see [34] for details on the calculation). In contrary, if one oscillator has its natural frequency located outside the excited band $[\omega_1, \omega_2]$ (i.e. if one oscillator is non resonant), approximation of energy flow between oscillators by equation (28) based on excitation in the entire frequency band $]-\infty, +\infty[$, is not correct. Indeed, the spectrum of energy flow exhibit significant values for frequencies close to the blocked natural frequency of oscillators, if one oscillator is non resonant in the band $[\omega_1, \omega_2]$, these significant values should not be taken into account for the evaluation of the energy flow. It is the reason why relation (28) (considering excitation in $]-\infty, +\infty[$) could not be used to estimate the power flow exchanged by oscillators if one (at least) is non resonant. Thus, the original SmEdA model [28] cannot describe correctly the energy flow between non resonant oscillators. Fig. 10 allows us to highlight this point on one example. The oscillator 1 is supposed to be excited by a white noise force in the third octave band 1000 Hz and to be coupled by a gyroscopic element ($\gamma = 0.002$) to oscillator 2. The blocked natural frequency of oscillator 1 is fixed to 1000 Hz (i.e. $\bar{\omega}_1 = 6283$ rad/s). The energy ratio of two coupled oscillators has been calculated by two methods; (a), a numerical resolution of Eq. (27) in the frequency domain that gives a reference result; (b), a SmEdA calculation using Eq. (28) and the energy balance of each oscillator (see [28]). The results are plotted in Fig. 10 in function of the natural frequency of oscillator 2. We can observe a good agreement of the two calculations when oscillator 2 is resonant. In contrary, if the natural frequency of oscillator 2 is outside the frequency band of excitation, significant discrepancies between the two calculations are observed. This result clearly shows that Eq. (28) and SmEdA model can not be used directly to estimate the power flow exchanged with non resonant modes. This explains why we have proposed the new modal scheme of Fig. 8 involving only couplings between resonant modes.

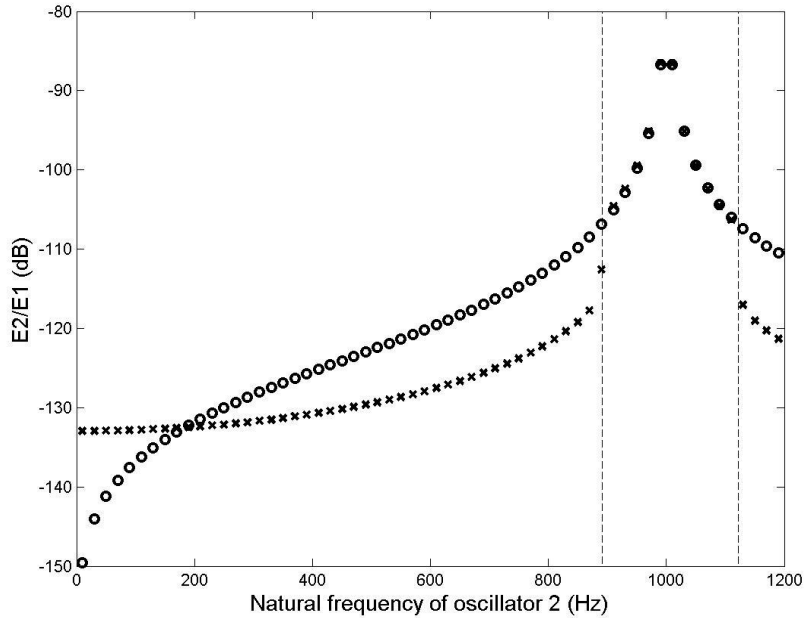

Figure 10. Oscillator energy ratio E2/E1 versus blocked natural frequency of oscillator 2. Two results: cross, reference; circle, SmEdA. Natural frequency of oscillator 1: 1000 Hz. Oscillator damping loss factors: $\eta_1$=0.01; $\eta_2$=0.001; Oscillator 1 excited by a white noise force in the third octave band of central frequency 1000 Hz (cut-off frequencies symbolised by vertical dashed line).

## *3.3 Energy equations of the cavity-structure-cavity system*

### 3.3.1 Modal energy equilibrium

Let us consider mode $p$ of excited cavity 1. The principle of energy conservation indicates that the injected power into mode $p$ by external excitation is either dissipated by internal damping of the mode or exchanged with modes of the other subsystems. In the present case, this can be written as:

$$\Pi_{inj}^{p} = \Pi_{diss}^{p} + \sum_{q \in \hat{Q}^R} \Pi_{pq} + \sum_{r \in \hat{R}} \Pi_{pr}, \quad \forall p \in \hat{P}, \tag{30}$$

where:
- $\Pi_{inj}^{p}$ is the time-averaged power injected by the generalized force $F_p$;
- $\Pi_{diss}^{p}$ is the time-averaged power dissipated by the internal damping of mode $p$;
- $\sum_{q \in Q^R} \Pi_{pq}$ is the time-averaged power flow exchanged by mode $p$ with the resonant plate modes;
- $\sum_{r \in \hat{R}} \Pi_{pr}$ is the time-averaged power flow exchanged by mode $p$ with the cavity 2 modes through the structure.

Let us estimate the different powers appearing in this equation:
- Evaluating $\Pi_{inj}^{p}$ from the injected power relation established for an oscillator excited by a white noise force ([28]) gives:

$$\Pi_{inj}^{p} = \frac{\pi}{4} \bar{S}_{Q_p}, \tag{31}$$

where $\bar{S}_{Q_p}$ is the power spectral density of the generalised source strength.

- The power dissipated by modal damping can be related to its total energy by:

$$\Pi_{diss}^{p} = \omega_p \eta_p E_p, \tag{32}$$

where $E_p$ is the time averaged energy of mode $p$, and $\eta_p$ is the modal damping factor.

- To evaluate the power exchanged by mode $p$ of cavity 1 with mode $q$ of the panel, we isolate these two modes in the modal equations (26):

$$\begin{cases} \varphi_p'' + \omega_p \eta_p \varphi_p' + \omega_p^2 \varphi_p - W_{pq} \chi_q' = L_{p-q}, \\ \chi_q'' + \omega_q \eta_q \chi_q' + \omega_q^2 \chi_q + W_{pq} \varphi_p' = L_{q-p}, \end{cases} \tag{33}$$

where:

$$L_{p-q} = Q_p + \sum_{\substack{\tilde{q} \in \hat{Q}^R \\ \tilde{q} \neq q}} W_{p\tilde{q}} \chi_{\tilde{q}}' + \sum_{r \in \hat{R}} \left( \sum_{q \in \hat{Q}^{NR}} W_{pq} W_{rq} \right) \chi_r, \quad L_{q-p} = \sum_{\substack{\tilde{p} \in \hat{P} \\ \tilde{p} \neq p}} W_{\tilde{p}q} \varphi_{\tilde{p}}' + \sum_{r \in \hat{R}} W_{rq} \zeta_r'. \tag{34}$$

Assuming - as done classically in SEA - that interaction forces $L_{p-q}$ and $L_{q-p}$ are uncorrelated white noise forces, the basic relation established for two coupled oscillators (28) can be used in the case of a gyroscopic coupling:

$$\Pi_{pq} \approx \beta_{pq} (E_p - E_q), \tag{35}$$

where $\beta_{pq}$ is the modal coupling factor. It is a function of natural angular frequencies, $\omega_p, \omega_q$; modal damping factors, $\eta_p, \eta_q$; and interaction modal works, $\mathbf{W}_{pq}$:

$$\beta_{pq} = (\mathbf{W}_{pq})^2 \left\{ \frac{\left( \omega_p \eta_p (\omega_q)^2 + \omega_q \eta_q (\omega_p)^2 \right)}{\left[ \left[ (\omega_p)^2 - (\omega_q)^2 \right]^2 + (\omega_p \eta_p + \omega_q \eta_q) \left[ \omega_p \eta_p (\omega_q)^2 + \omega_q \eta_q (\omega_p)^2 \right] \right]} \right\}. \tag{36}$$

- With the same process, we can estimate the power exchanged by mode $p$ of cavity 1 with mode $r$ of cavity 2, by isolating these two modes in the modal equations (26):

$$\begin{cases} \varphi_p'' + \omega_p \eta_p \varphi_p' + \omega_p^2 \varphi_p - \left( \sum_{q \in \hat{Q}_{nr}} W_{pq} W_{rq} \right) \zeta_r = L_{p-r} \\ \zeta_r'' + \omega_r \eta_r \zeta_r' + \omega_r^2 \zeta_r - \left( \sum_{q \in \hat{Q}_{nr}} W_{pq} W_{rq} \right) \varphi_p = L_{r-p} \end{cases} \tag{37}$$

where

$$L_{p-r} = Q_p + \sum_{q \in \hat{Q}^R} W_{pq} \chi_q' + \sum_{\substack{\tilde{r} \in \hat{R} \\ \tilde{r} \neq r}} \left( \sum_{q \in Q^{NR}} W_{pq} W_{\tilde{r}q} \right) \chi_{\tilde{r}}, \quad L_{r-p} = \sum_{q \in \hat{Q}^R} W_{rq} \chi_q' + \sum_{\substack{\tilde{p} \in \hat{P} \\ \tilde{p} \neq p}} \left( \sum_{q \in Q^{NR}} W_{\tilde{p}q} W_{rq} \right) \varphi_{\tilde{p}}. \tag{38}$$

Again, assuming that interaction forces $L_{p-r}$ and $L_{r-p}$ are uncorrelated white noise forces, the relation (28) for oscillators coupled by a spring can be used:

$$\Pi_{pr} \approx \beta_{pr} (E_p - E_r), \tag{39}$$

where $\beta_{pr}$ is the modal coupling factor given by:

$$\beta_{pr} = \left( \sum_{q \in Q^{NR}} W_{pq} W_{rq} \right)^2 \left\{ \frac{(\omega_p \eta_p + \omega_r \eta_r)}{\left[ (\omega_p)^2 - (\omega_r)^2 \right]^2 + (\omega_p \eta_p + \omega_r \eta_r) \left[ \omega_p \eta_p (\omega_r)^2 + \omega_r \eta_r (\omega_p)^2 \right]} \right\}. \tag{40}$$

By introducing (31, 32, 35, 39) in (30), we obtain the power balance equation for mode $p$ of cavity 1:

$$\Pi_{inj}^p = \omega_p \eta_p E_p + \sum_{q \in \hat{Q}^R} \beta_{pq} (E_p - E_q) + \sum_{r \in \hat{R}} \beta_{pr} (E_p - E_r), \quad \forall p \in \hat{P}. \tag{41}$$

The power balance equation for the resonant modes of the plate and cavity 2 can be written in the same way (considering that these subsystems are not directly excited by the external source). Finally, we obtain a linear equation system having the modal energy as unknowns:

$$\begin{cases} \omega_p \eta_p E_p + \sum_{q \in \hat{Q}^R} \beta_{pq} (E_p - E_q) + \sum_{r \in \hat{R}} \beta_{pr} (E_p - E_r) = \Pi_{inj}^p, & \forall p \in \hat{P}, \\ \omega_q \eta_q E_q + \sum_{p \in \hat{P}} \beta_{pq} (E_q - E_p) + \sum_{r \in \hat{R}} \beta_{qr} (E_q - E_r) = 0, & \forall q \in \hat{Q}^R, \\ \omega_r \eta_r E_r + \sum_{q \in \hat{Q}^R} \beta_{qr} (E_r - E_q) + \sum_{p \in \hat{P}} \beta_{pr} (E_r - E_p) = 0, & \forall r \in \hat{R}. \end{cases} \tag{42}$$

This equation system can be solved and the total energy of each cavity can finally be obtained by adding modal energies:

$$E_{C1} = \sum_{p \in \hat{P}} E_p, \quad E_{C2} = \sum_{r \in \hat{R}} E_r, \tag{43}$$

where $E_{C1}$ (resp. $E_{C2}$) is the time-averaged total energy of cavity 1 (resp. cavity 2). The energy noise reduction (ENR) or the transmission loss can then be deduced using (2).

(42) consists in the energy equations of the SmEdA model including the non resonant transmission though the structure. In the next section, we use this model to deduce an SEA model including the non resonant transmission.

### 3.3.2 Modal energy equipartition

In classical SEA, modal energy equipartition is assumed and permits restricting the $P + Q_r + R$ degrees of freedom (DOF) of the SmEdA model (i.e. (42)) to only three DOF, one per subsystem. Considering that for the frequency band of angular central frequency, we can approximate the modal damping bandwidths by,

$$\omega_p \eta_p \approx \omega_c \eta_{C1}, \ \forall p \in \hat{P}, \ \omega_q \eta_q \approx \omega_c \eta_S, \ \forall q \in \hat{Q}^R, \ \omega_r \eta_r \approx \omega_c \eta_{C2}, \ \forall r \in \hat{R}, \text{ and,} \quad (44)$$

by introducing equipartition relation (45),

$$E_p = \frac{E_{C1}}{P}, \ \forall p \in \hat{P}, \ E_q = \frac{E_S}{Q^R}, \ \forall q \in \hat{Q}_r, \ E_r = \frac{E_{C2}}{R}, \ \forall r \in \hat{R} \quad (45)$$

in the modal energy equations (42), and adding the modal equations of each subsystem, we obtain the standard SEA equation:

$$\begin{cases} \Pi_{inj}^1 = \omega_c \eta_{C1} E_{C1} + \omega_c \eta_{C1-S}\left(E_{C1} - \frac{P}{Q_r} E_S\right) + \omega_c \eta_{C1-C2}\left(E_{C1} - \frac{P}{R} E_{C2}\right), \\ 0 = \omega_c \eta_S E_S + \omega_c \eta_{C1-S}\left(\frac{P}{Q_r} E_S - E_{C1}\right) + \omega_c \eta_{C2-S}\left(\frac{R}{Q_r} E_S - E_{C2}\right), \\ 0 = \omega_c \eta_{C2} E_{C2} + \omega_c \eta_{C2-S}\left(E_{C2} - \frac{R}{Q_r} E_P\right) + \omega_c \eta_{C1-C2}\left(\frac{P}{R} E_{C2} - E_{C1}\right), \end{cases} \quad (46)$$

where $E_S$ is the total energy of the structure, $\Pi_{inj}^1 = \sum_{p \in \hat{P}} \Pi_{inj}^p$ represents the power injected by external sources in cavity 1 and the SEA coupling loss factors are given by:

$$\eta_{C1-S} = \frac{1}{P\omega_c} \sum_{p \in \hat{P}} \sum_{q \in \hat{Q}^R} \beta_{pq}, \ \eta_{S-C2} = \frac{1}{Q^R \omega_c} \sum_{q \in \hat{Q}^R} \sum_{r \in \hat{R}} \beta_{qr}, \text{ and,} \quad (47)$$

$$\eta_{C1-C2} = \frac{1}{P\omega_c} \sum_{p \in \hat{P}} \sum_{r \in \hat{R}} \beta_{pr}. \quad (48)$$

$\eta_{C1-C2}$ is the coupling loss factor between the 2 cavities used to describe the non-resonant transmission in the SEA model. Previously, a coupling loss factor between the two cavities was introduced by Crocker and Price [2] in the SEA model to simulate the mass law transmission of the panels. However, this factor was introduced without justification of the basic formulation of the SEA model dedicated to describing the resonant transmission. The present development provides the justification of this coupling loss factor between the two cavities that are not directly connected. Moreover, an expression of this parameter (48) as a function of the modal information (modal frequency, mode shapes) is obtained for each subsystem. The complex geometries and mechanical properties of the structure and the cavity can then be easily taken into account. Of course, this is true provided that the subsystems can be modelled by Finite Element Modelling and that the modal information of each subsystem can be extracted with the available computation resources. The application of this approach has an upper frequency bound. As the modal FEM calculations must be performed for each independent subsystem and not for the global system, this frequency bound could be relatively high. An example of the application is proposed in section V.

### 3.3.3 Simplified energy expressions

Two paths of energy transmission were identified between the two cavities: the resonant transmission corresponding to the coupling of the cavity modes through the resonant structure modes and the non-resonant transmission which are simulated by the direct modal coupling of the two cavities. Since the coupling between the three subsystems can be qualified as weak (due to high impedance ruptures between the air cavity and the structure), the modal coupling factors are therefore significantly lower than the modal damping bandwidth. Consequently, the modal energies are significantly different from one subsystem to another. In general, we obtain:

$$\beta_{pq} \square \ \omega_p \eta_p, \ \beta_{qr} \square \ \omega_q \eta_q, \ \beta_{pr} \square \ \omega_r \eta_r,$$
$$E_r \square \ E_q \square \ E_p, \ \forall p \in \hat{P}, \ \forall q \in \hat{Q}^R, \ \forall r \in \hat{R}. \quad (49)$$

Under these conditions, the energy equation (42) can be simplified and the modal energy of the receiving cavity estimated for each transmission path. Indeed, if the non resonant transmission path in (42) and the conditions (49) are considered, we can write:

$$\begin{cases} \omega_p \eta_p E_p \approx \Pi_{inj}^p, & \forall p \in \hat{P}, \\ \omega_r \eta_r E_r^{NR} - \sum_{p \in \hat{P}} \beta_{pr} E_p \approx 0, & \forall r \in \hat{R}, \end{cases} \quad (50)$$

where $E_r^{NR}$ is the modal energy of the receiving cavity due to the non resonant transmission. The following can be deduced:

$$E_p \approx \frac{\Pi_{inj}^p}{\omega_p \eta_p}, \ \forall p \in \hat{P}, \ \text{and,} \ E_r^{NR} \approx \frac{1}{\omega_r \eta_r} \sum_{p \in \hat{P}} \beta_{pr} \frac{\Pi_{inj}^p}{\omega_p \eta_p}, \quad \forall r \in \hat{R}. \quad (51)$$

With the same process, an approximation of the modal energy of the receiving cavity can be given using the resonant transmission, $E_r^R$ :

$$E_r^R \approx \frac{1}{\omega_r \eta_r} \sum_{p \in \hat{P}} \beta'_{pr} \frac{\Pi_{inj}^p}{\omega_p \eta_p}, \quad \forall r \in \hat{R} \quad (52)$$

with $\beta'_{pr} = \sum_{q \in \hat{Q}^R} \left( \frac{\beta_{qr} \beta_{pq}}{\omega_q \eta_q} \right).$ (53)

Still under the conditions of (49), an approximation of the modal energy of the receiving cavity due to the two path transmissions can be obtained by:

$$E_r \approx E_r^R + E_r^{NR}, \quad \forall r \in \hat{R} \quad (54)$$

(51-53) allow us to estimate the modal energies of each subsystem with a very short computation time despite the large number of modes. Moreover, these latter results highlight the modal parameters which play a significant role in each transmission path, as is shown in the next section.

# 4. Application to the cavity-plate-cavity system

In this section, we apply the SmEdA model to the cavity-plate-cavity system described in section 2.2. The numerical data and the source position remain unchanged.

## 4.1. Modal energy distribution for Resonant and Non-Resonant path transfer

In section 3.3.3, we proposed simplified energy balance relations which give us the analytical expression of the modal energies resulting of the Resonant and Non-Resonant transmission path. In Fig. 11, we compare these approximate modal energies for the receiving cavity with those obtained with the full energy balance relations of section 3.3.1. (i.e. Eq. (42)). Good agreement can be seen between the two calculations, thereby validating the assumptions of the simplified model for this case. The accuracy obtained is slightly less good for the Resonant path compared to the Non-Resonant path. This can be explained by a cumulative error for the Resonant path. Indeed, for the Resonant path the energy is shared between the path from the excited cavity to the plate and the path from the plate to the receiving cavity, whereas the energy is shared directly between the excited cavity and the receiving cavity for the Non-Resonant path (in the present energy model). However, the differences are not significant.

On the other hand, it can be seen in Fig. 11 that the modal energies are not uniformly distributed, contrary to the classical SEA assumption. If the assumption of modal energy equipartition is considered (i.e. Eq. (45)), we again obtain an Energy Noise Reduction (ENR) of 40.5 dB and 44.5 dB without this assumption for the Resonant path. For the Non-Resonant path, the difference is less considerable, 31.3 dB versus 29.1 dB. The significant variations of the modal energy distribution can be explained, on the one hand, by a relatively low modal overlap at this frequency (around 1); and on the other hand, by the fact that few modes effectively participate in the coupling due to the spatial coincidence. This effect can be highlighted by the modal coupling factors as proposed in matrix form in Fig. 12. The highest values of these parameters are contained on the pseudo-diagonal of the matrix. This is due to the coincidence frequency phenomena as discussed and called "maximum proximate modal coupling" by Fahy [8]. Using the expression (36) of the modal coupling factors, $\beta_{pq}$, we can write the approximation [8]:

$$\beta_{pq} \approx \begin{cases} \dfrac{(W_{pq})^2}{(\omega_p \eta_p + \omega_q \eta_q)} & \text{if } 2|\omega_p - \omega_q| < (\omega_p \eta_p + \omega_q \eta_q) \\ 0 & \text{if } 2|\omega_p - \omega_q| > (\omega_p \eta_p + \omega_q \eta_q) \end{cases} \quad (55)$$

For the maximum proximate modal couplings, $\beta_{pq}$ depends on the modal interaction work, $W_{pq}$. This parameter expresses the strength of the spatial coincidence between the subsystem modes and can vary strongly from one couple of modes to another as could be observed previously in Fig. 5. The result is that few couples of modes effectively participate in the coupling between the excited cavity and the plate (as see in Fig. 12a). The same behaviour is observed for the coupling between the plate and the receiving cavity (see Fig. 12b) and for the non-resonant path, between the excited cavity and the receiving cavity (see Fig. 12c). It may be surprising at first sight that the values of the modal coupling factors of Fig. 12c (which represent the NR path) are not greater than those of Figs. 12a and 12b, whereas the energy level of the receiving cavity due to the NR path is significantly greater that that of the R path (as observed in Fig. 11). It should be remembered that the NR path is represented by the direct coupling of the cavity modes whereas these modes are coupled by the intermediary of the plate modes for the R path. The modal coupling factors cannot therefore be compared directly. It is more relevant to consider the equivalent modal coupling factors of the simplified model (53) for the R path, as proposed in Fig. 12d. It can be seen that the maximum values of this parameter are ten times lower than those of Fig. 12c. This explains why, in our model, we obtained an energy level 10 dB higher for the NR path than for the R path.

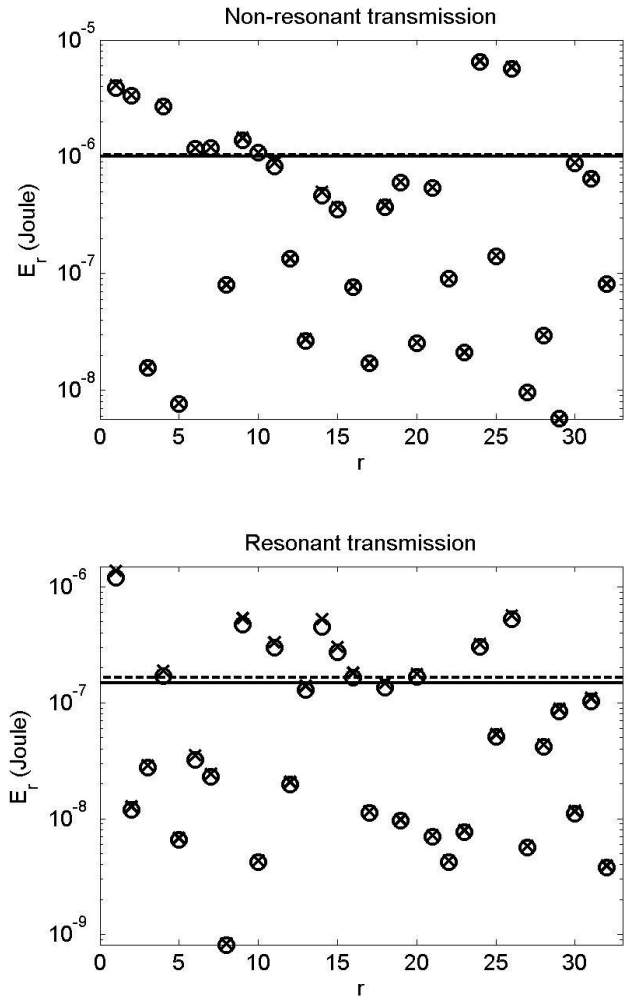

Figure 11. Modal energy distribution of the receiving cavity for the third octave band 1000 Hz. (a), considering only the non-resonant transmission; (b), considering only the resonant transmission. Circle, full model (i.e. (42)); cross, simplified model (i.e. (51-53)). Modal energy averaged: solid line, full model; dash line, simplified model.

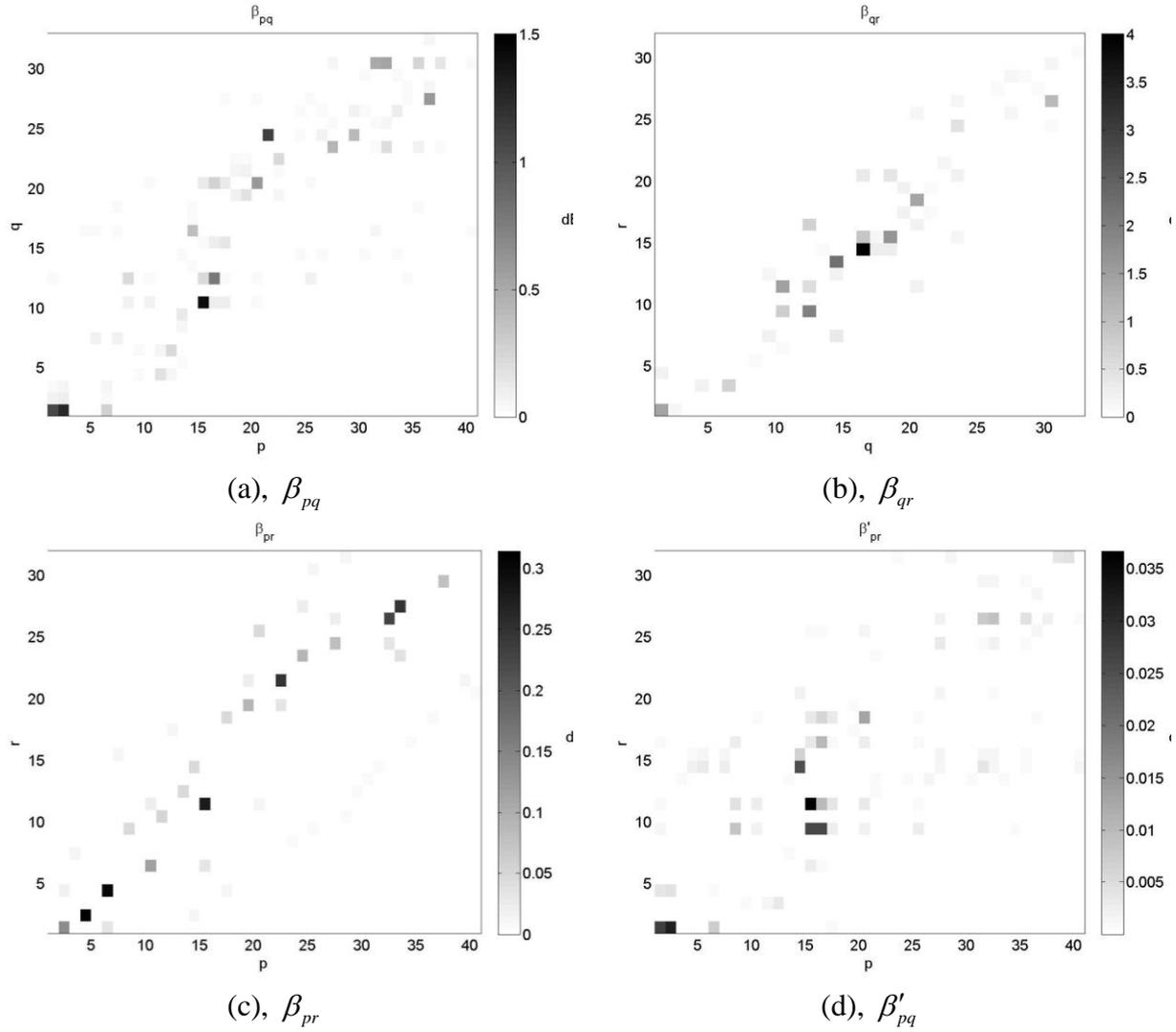

Figure 12. Modal coupling factors between: (a), cavity 1 and the plate; (b), the plate and cavity 2; (c), cavity 1 and cavity 2 representing the non resonant transmission; (d), cavity 1 and cavity 2 representing the resonant transmission. Results for the third octave band 1000 Hz.

## 4.2. Comparison with DMF results

For the purposes of validation, in Fig. 13 we compare the ENR predicted by SmEdA (51-53) and DMF (14-18) for two calculations: one taking the NR modes into account and the other one considering only the R modes. Generally good agreement between the two models can be observed for both the R and NR paths. Some discrepancies of 2 or 3 dB can be observed in certain frequency bands. In general, SmEdA slightly underestimates the energy transmission, whereas it slightly overestimates it for some frequency bands around the critical frequency. These discrepancies can be related to the approximation made for estimating the power flow between two coupled modes from the relation established for two coupled oscillators (i.e. 35, 39). The interaction forces $L_{p-q}$ and $L_{q-p}$ (resp. $L_{p-r}$ and $L_{r-p}$) do not fully conform to the assumption of uncorrelated white noise forces. We emphasize that this assumption is also considered in the SEA model. Considering wave-mode duality, it is well-known that regarding wave propagations, this assumption implies neglecting the direct field (i.e. coherent part) of the vibratory field of each subsystem compared to the reverberant field (i.e. incoherent part) [35].

The results of Fig. 13 remain fully satisfactory and they clearly show that the extension of SmEdA presented in this paper is relevant.

We can emphasize that SmEdA permit to reduce significantly the computing time compared to a DMF calculation. Indeed, SmEdA gives us directly the energy per third octave band from Eq. (51-53) whereas DMF necessitate to discretize the frequency band with a frequency resolution depending on the damping bandwidth, to resolving Eq. (16-18) (with a matrix inversion) for each of these frequencies and to sum the frequency energy results to deduce the energy per third octave band. For indication, the DMF result of Fig. 13 for the third octave 5 kHz has been obtained in 925 seconds whereas only 29 seconds have been required by SmEdA (calculations achieved with MATLAB on a Personal Computer Intel Xeon 2.67 GHz).

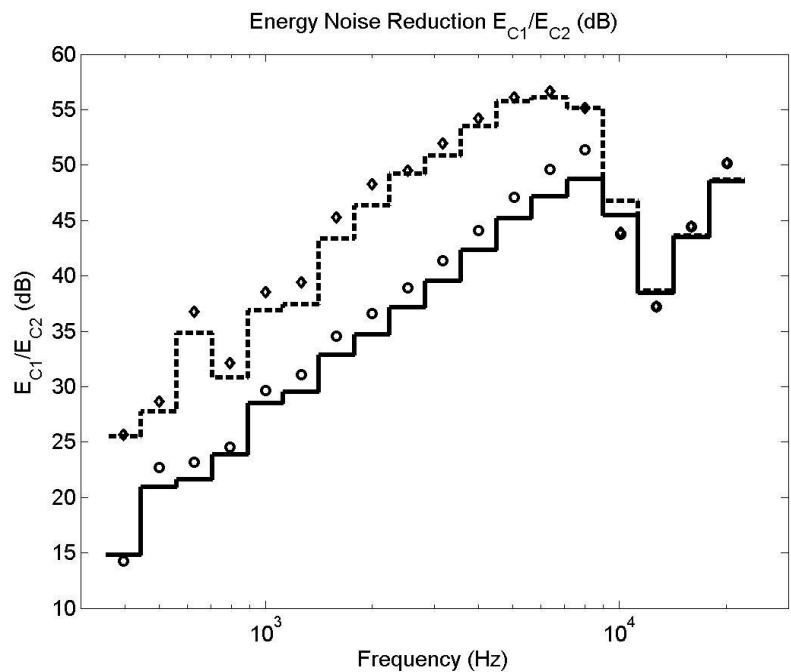

Figure 13. ENR versus third octave band. Comparison of four calculations: solid line, DMF with NR plate modes; circles, SmEdA taking the NR plate modes into account; dashed line, DMF without NR plate modes; diamonds, SmEdA without NR plate modes.

### *4.3. Influence of the plate damping*

It is well known that for infinite plates below the critical frequency, the Transmission Loss is controlled by the mass law of the plate and is independent of the plate damping. For finite plates coupled to closed cavities, the cavity modes can play a significant role and two paths of transmission exist: Resonant and Non-Resonant. If the Non-Resonant path dominates (as in the case of Fig. 13), the plate damping should not influence the TL. On the contrary, if the Resonant path dominates, the TL should decrease when the plate damping increases.

For the present case, Fig. 14 shows the influence of the plate damping as a function of the frequency. The calculations are performed with the SmEdA model including the NR plate modes. It can be seen that for frequencies well below the critical frequency, the ENR is practically unchanged when the plate Damping Loss Factor (DLF) varies from 0.1 to 0.01, whereas small variations (around 2-3 dB) can be observed when it varies from 0.01 to 0.001. The higher the damping factor of the plate, the more highly Non-Resonant the transmission becomes. When the DLF is around 0.001, the Resonant path cannot be neglected which explains why a small variation of the ENR can be seen below the critical frequency. The influence of the plate damping can be analysed in the simplified energy equations (52, 53). Relation (52) indicates that the energy transmission by the NR path is independent on the plate DLF whereas it is dependent for the R path regarding Eq. (52, 53) (i.e. plate DLF in the denominator of $\beta'_{pr}$). Then, when the plate DLF decreases, the energy transmission by the NR path is unchanged and the energy transmission by the R path increases. The result is that for a given plate DLF, the R path can become dominant.

For frequencies above the critical frequency, the R path is generally dominant due to the space-frequency coincidences of certain couples of modes. The plate DLF therefore directly influences the TL, as observed in Fig. 14.

In conclusion, the present SmEdA model is able to represent the plate damping effect: if the NR path dominates (as in the case of an infinite plate), the plate damping does not influence the TL. On the contrary, if the R path dominates, the TL decreases when the plate DLF increases. The latter case occurs for frequencies above the critical frequency or when the plate and the cavity are weakly damped.

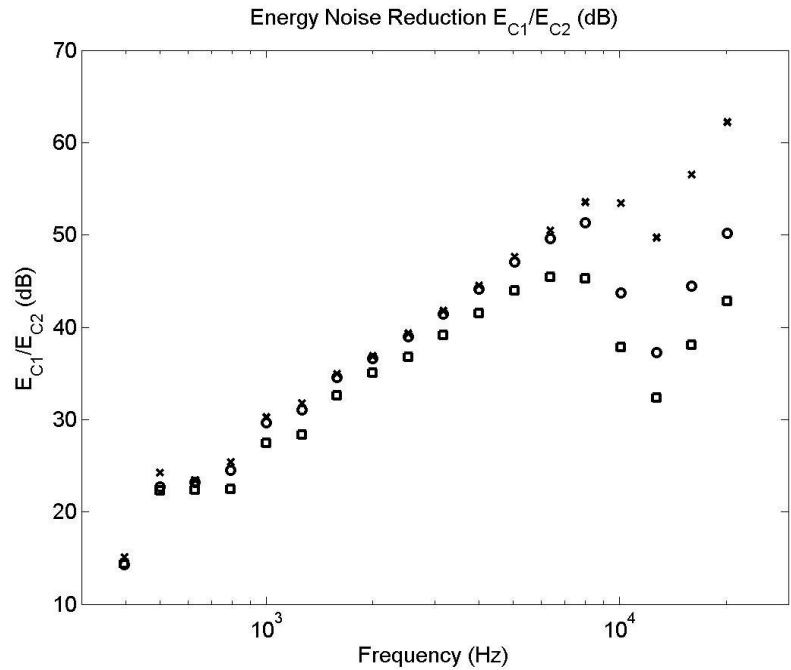

Figure 14. ENR versus third octave band for three plate Damping Loss Factor: cross, 10% (i.e. $\eta_S = 0.1$). circles, 1% (i.e. $\eta_S = 0.01$) ; square, 0.1% (i.e. $\eta_S = 0.001$). SmEdA results. $\eta_{C1} = \eta_{C2} = 0.01$.

# 5. Illustration of TL estimation for a complex structure

The SmEdA model presented in this paper is based on prior knowledge of the subsystem modes. For each uncoupled subsystem, the natural frequency and the mode shapes on the coupling boundary should be estimated. The subsystem modes can be calculated using Finite Element Models (FEM) for cavities with complex geometries or for structures with complex geometries or mechanical properties. In this section, we propose to estimate the TL of a ribbed plate in order to illustrate the calculation process mixing SmEdA and FEM.

Let us consider the previous rectangular plate (i.e. 0.8 m x 0.6m, 1mm thick) stiffened by ribs regularly spaced along its longer edge. The rib cross-section is a 5mm x 5mm square and the rib spacing is 50 mm. The ribs and the plate are made of steel (see mechanical characteristics in section 2.2) and the plate is assumed to be simply-supported at its four edges. To evaluate the TL of this stiffened plate, we consider the parallelepiped cavities of the test case (same dimensions and fluid properties). The modal information of the cavities is still calculated analytically. With the present approach, cavities with complex geometries could be considered for studying the effect of the geometry or the source location on the TL. Of course, as the modal density of a cavity increases with the square of the frequency, the amount of numerical data calculated by FEM (i.e. mode shapes) could become dramatically huge when the frequency increases. This is not a major drawback, however, as it is well known that the geometry of the cavity essentially has an effect on the TL at low-frequency [4].

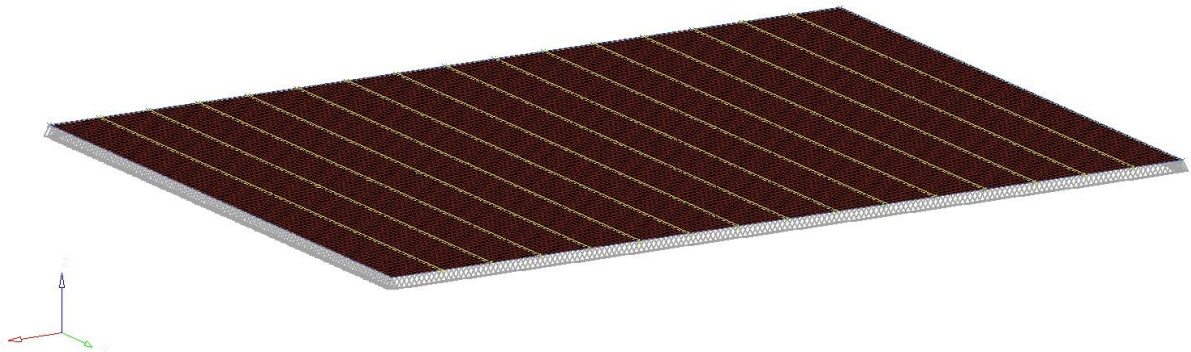

Figure 15. Finite element meshing of the ribbed plate.
MD NASTRAN model: 19481 Nodes, 19200 CQUAD4, 1800 CBEAM.

A Finite Element Model of the ribbed plate was built. The plate is modelled with 2D shell elements whereas the ribs are modelled with 1D beam elements. A criterion of six elements by flexural wavelength at 10 kHz is considered. The FE model is composed of 19481 nodes, 19200 2-D elements (CQUAD4), and 1800 1-D elements (CBEAM) as shown on Fig. 15. The normal mode analysis (SOL103) was performed with the MD NASTRAN solver on a standard Personal Computer (Intel Xeon 2.67 GHz). 1359 modes below 10 kHz were extracted in 10 minutes. The natural frequencies and the mode shapes were saved in a PCH file of 4.2 GBytes. This file was imported on MATLAB to perform the SmEdA calculation. The modal interaction works (8) were then approximated by estimating the integral with the rectangular rule:

$$W_{pq} = \frac{S}{N} \sum_{i=1}^{N} \tilde{W}_q^i \tilde{p}_p^i, \quad W_{qr} = \frac{S}{N} \sum_{i=1}^{N} \tilde{W}_q^i \tilde{p}_r^i, \tag{56}$$

where:    - $\tilde{W}_q^i$ is the plate normal displacement for the $q^{th}$ mode at node $i$ (NASTRAN results);

- $\tilde{p}_p^i$ (resp. $\tilde{p}_r^i$) is the cavity pressure for the $p^{th}$ mode (resp. $r^{th}$ mode) calculated at the position of node $i$;

- $N$ is the node number and $S$ is the plate area.

In order to validate the numerical process and to study the effect of the numerical errors introduced by FE discretisation, the calculations were also performed by considering an unribbed plate. The SmEdA results considering mode information calculated analytically and numerically are compared in Fig. 16 for the bare plate. Good agreement can be observed throughout the frequency band. Although some discrepancies can be noticed between the modal frequencies calculated analytically and numerically (up to 5% for the highest frequencies), the SmEdA results are not sensitive to them. This is certainly due to a data averaging effect, as observed previously in [28] for two coupled plates.

The SmEdA calculation was performed for the ribbed plate. The most time-consuming part of the process consists in the evaluation of the modal interaction work for each mode couple with (56). This task is currently done with MATLAB (with different loops). It could be optimised in the future by using a C or FORTRAN code. The results up to the third octave band 8 kHz were obtained in 4 hours on the PC described previously.

The ENR of the ribbed plate is plotted in Fig. 17. By comparing it with Fig. 16, it can be seen that the ENR is significantly influenced by the presence of the ribs in agreement with the literature [18, 37]. Moreover, the difference between the SmEdA results with and without the NR plate modes varies with the frequency and it is smaller than for the bare plate (see Fig. 13). This indicates that the R path plays a bigger role for the ribbed plate than for the bare plate at these frequencies and for the data considered. This can be explained by the fact that the ribs increase the stiffness of the plate in one direction which may be considered as orthotropic (when the rib spacing is less than about a third of the flexural wavelength of the plate [36]). Although there is just one critical frequency for the bare plate (i.e. around 11 kHz), for the equivalent orthotropic plate, the critical frequency is dependent on the direction of the incident acoustic wave. The lower critical frequency corresponds to a wave travelling in the plate's stiffest direction [38]. Using the relations defining the characteristics of the equivalent orthotropic plate given in [36], this lower frequency is evaluated at 4 kHz for the present case. The results given in Fig. 17 are in full agreement with this phenomenon. As the ribbed plate was simply chosen as an example to illustrate the calculation process described in this paper, the behaviour of this ribbed plate is not subject to detailed study here.

The sound transmission of a ribbed plate was evaluated with the present approach in satisfactory computing time and this approach can be used to evaluate the TL of other complex structures such as a car firewall or a truck floor. Moreover, the geometry of excited and receiving cavities such as the engine and passenger compartments for automotive applications can be taken into account.

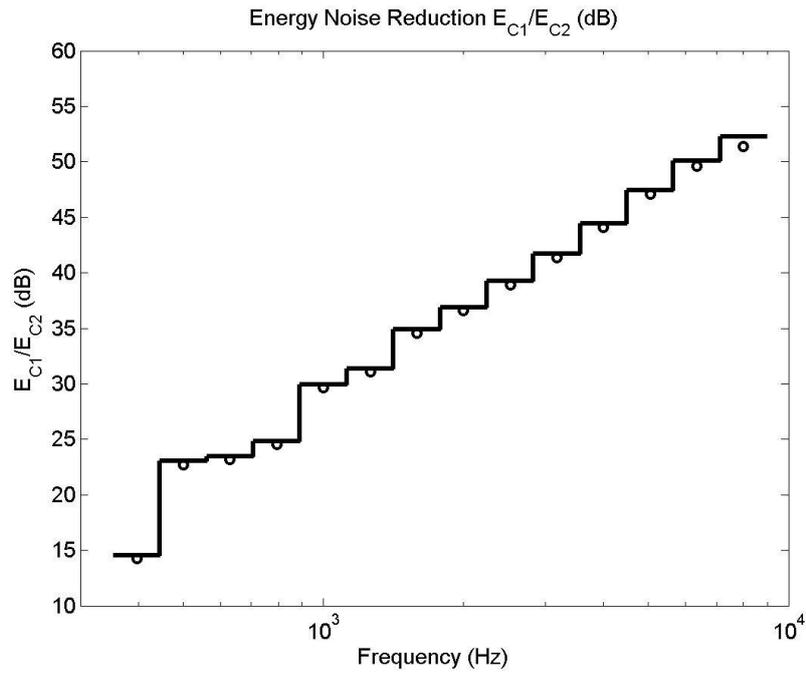

Figure 16. ENR versus third octave band for the bare plate. Comparison of the SmEdA results with plate modes calculated analytically (circle) and numerically with FEM (solid line).

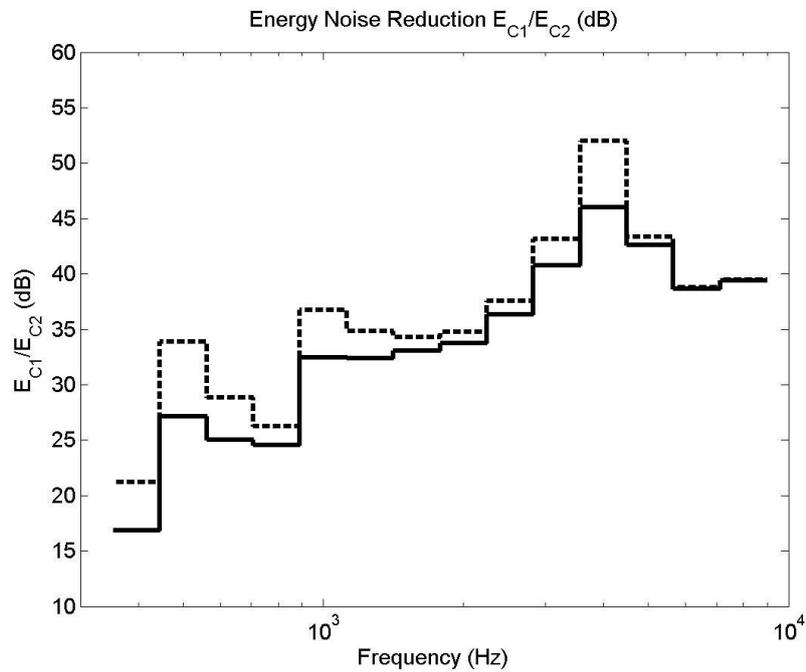

Figure 17. ENR versus third octave band for the ribbed plate: solid line, SmEdA results with the NR plate modes; dashed line, SmEdA results without the NR plate modes.

# 6. Conclusions

An extension of SmEdA taking the Non Resonant mode contributions into account was presented in order to estimate the TL of a structure between two cavities. The developments are based on the condensation of the NR modes in the modal equations. They led to a new modal coupling scheme describing the behaviour of the cavity-structure-cavity system: the Resonant path was described by the gyroscopic couplings of the structure and cavity modes whereas the Non-Resonant path was represented by the stiffness couplings of the modes of the two cavities. The energy equations of SmEdA were fully defined using the source characteristics (i.e. position, level) and the modal information of each uncoupled subsystem (natural frequencies, mode shapes on the coupling boundary and the possible excitation position). The TL of the structure can be easily deduced by resolving these equations for a given frequency band.

Moreover, the introduction of the modal energy equipartition assumption in the energy equations led to the SEA equations. A Coupling Loss Factor between the two cavities appeared in the equations providing proof of the factor introduced long ago by Crocker and Price [2] when considering the TL of infinite panels. The expression of this CLF depends on the shape and frequency of the R cavity modes and the NR structure modes. It can be applied to various sorts of finite panels, contrary to the expression given in [2] which is adapted for an infinite flat panel.

Comparisons between SmEdA and DMF results allowed us to validate the developments presented. SmEdA calculations for different plate dampings also highlighted variations of transmission loss in agreement with expectations as a function of the dominant path (i.e. R or NR).

The proposed process can be easily used to evaluate the TL of complex structures, as was illustrated for a ribbed plate. The information of the subsystem mode can then be calculated using Finite Element Models. In addition, cavities with complex geometries and structures with complex geometries and various mechanical properties can be considered.

In the future, the process could be improved by taking into account the effect of trims on the structure and absorbing materials inside the cavities. This would require evaluating the modal Damping Loss Factors from Finite Element Models including trims and absorbing materials. This methodology is briefly described in [39] and it will be the subject of a future paper.

# Acknowledgments

This work was funded jointly by the French Government (FUI 12 - Fonds Unique Interministériel) and the European Union (FEDER - Fonds européen de développement régional ). It was carried out in the framework of the LabEx CeLyA ("Centre Lyonnais d'Acoustique", ANR-10-LABX-60) and the research project CLIC ("City Lightweight Innovative Cab") bearing the label of the LUTB cluster (Lyon Urban Truck and Bus).